\documentclass[12pt, draftclsnofoot, journal, letterpaper, onecolumn]{IEEEtran}

\makeatletter
\def\ps@headings{%
\def\@oddhead{\mbox{}\scriptsize\rightmark \hfil \thepage}%
\def\@evenhead{\scriptsize\thepage \hfil \leftmark\mbox{}}%
\def\@oddfoot{}%
\def\@evenfoot{}}
\makeatother 

\IEEEoverridecommandlockouts

\usepackage{amsfonts}
\usepackage[dvips]{graphicx}
\usepackage{caption}
\usepackage{subfigure}
\usepackage{times}
\usepackage{cite}
\usepackage{lettrine}
\usepackage{amsmath}
\usepackage{amsmath}
\allowdisplaybreaks[4]
\usepackage{array}
\usepackage{amssymb}

\usepackage{stfloats}
\usepackage{slashbox}
\usepackage{graphicx}
\usepackage{footnote}
\usepackage{booktabs}
\usepackage{array}
\usepackage{algorithmic}
\usepackage{algorithm}
\usepackage{subeqnarray}
\usepackage{cases}
\usepackage{threeparttable}
\usepackage{color}
\usepackage{bm}

\UseRawInputEncoding
\begin{document}

\title{Bayesian Optimization for Online Management in Dynamic Mobile Edge Computing }

\IEEEoverridecommandlockouts

\author{Jia~Yan,~\IEEEmembership{Member,~IEEE}, Qin~Lu,~\IEEEmembership{Member,~IEEE}, and Georgios~B.~Giannakis,~\IEEEmembership{Fellow,~IEEE}
\thanks{Part of this work will be presented at the IEEE 56th Asilomar Conference on Signals, Systems, and Computers, Pacific Grove, California, USA, October 30th - November 2nd, 2022 \cite{jiaasilomar}. J. Yan, Q. Lu and G. B. Giannakis are with the Department of Electrical and Computer Engineering, University of Minnesota, Twin Cities, MN, USA. Emails: \{yanj,qlu,georgios\}@umn.edu.
This work was supported by NSF grants 1901134, 2126052, and 2128593.}
}

\maketitle

\vspace{-1.5cm}

\begin{abstract}
   Recent years have witnessed the emergence of mobile edge computing (MEC), on the premise of a cost-effective enhancement in the computational ability of hardware-constrained wireless devices (WDs) comprising the Internet of Things (IoT). In a general multi-server multi-user MEC system, each WD has a computational task to execute and has to select binary (off)loading decisions, along with the analog-amplitude resource allocation variables in an online manner, with the goal of minimizing the overall energy-delay cost (EDC) with dynamic system states. While past works typically rely on the explicit expression of the EDC function, the present contribution considers a practical setting, where in lieu of system state information, the EDC function is not available in analytical form, and instead only the function values at queried points are revealed. Towards tackling such a challenging online combinatorial problem with only bandit information, novel Bayesian optimization (BO) based approaches are put forth by leveraging the multi-armed bandit (MAB) framework. Per time slot, the discrete offloading decisions are first obtained via the MAB method, and the analog resource allocation variables are subsequently optimized using the BO selection rule. By exploiting both temporal and contextual information, two novel BO approaches, termed time-varying BO and contextual time-varying BO, are developed.  Numerical tests validate the merits of the proposed BO approaches compared with contemporary benchmarks under different MEC network sizes.

\end{abstract}

\section{Introduction}

The era of massive connectivity is brought into being by the Internet of Things (IoT), where tens of billions of wireless devices (WDs) are ubiquitously connected to the Internet through cellular networks. Constrained by limited batteries and low-power on-chip computing units, the WDs face challenges to support latency-sensitive applications in the current IoT paradigms such as autonomous driving, online gaming and virtual reality. To meet the intensive computation demands far beyond the WDs' capacities, mobile edge computing (MEC) has emerged as a promising technology by releasing and distributing computing resources to the edge servers within the radio access networks to facilitate real-time services. Capitalizing on the MEC architecture, WDs in the IoT are able to carry out high-performance computation by offloading tasks to the servers located at the network edge \cite{MECsurvey1}. Compared with traditional mobile cloud computing, the MEC no longer suffers from high overhead and long backhaul latency.

Due to the time-varying wireless channel conditions and the heterogeneity in both the WDs and edge servers, judiciously offloading computations can offer significant performance enhancement. In general, MEC has two computation offloading models, referred to as binary and partial offloading \cite{MECsurvey1}. Binary offloading requires each task to be either executed locally or offloaded to the edge server as a whole \cite{MEC2}. On the other hand, a task under partial offloading model is allowed to be partitioned and computed both locally and at the edge server \cite{MEC1,wang2016mobile}. In this work, we focus on binary computation task offloading, which is commonly used in IoT to process indivisible simple tasks such as face recognition and temperature monitoring in smart home \cite{MECsurvey1}.    Prior works on offloading computations typically focus on offline algorithms by adopting either convex \cite{MEC1,MEC2} or non-convex (e.g., convex relaxation \cite{MEC3} and  heuristic local search \cite{bi2018computation,yan2019optimal}) optimization methods, which assume that the system states are known a priori, even though such knowledge is challenging to acquire beforehand.

With unknown system dynamics, online computational task offloading approaches have been extensively investigated. Building on the assumption of stationarity, a class of online algorithms rely on stochastic optimization methods such as Lyapunov optimization to determine the task offloading decisions within each time slot without future information \cite{mao2016dynamic,mao2017stochastic,yang2022dynamic}. Nevertheless, the nonstationarity introduced by the human participation in IoT makes the stochastic optimization impractical. Targeting at the nonstationary system dynamics, existing works focus on the online convex optimization (OCO) algorithms \cite{chen2018heterogeneous,chen2017online,hall2015online}, where the sequence of convex task offloading costs changes in an unknown and possibly adversarial manner. Yet, the OCO approaches necessitate the availability of explicit cost function forms or their gradients.

In practice though, the unpredictable WD preferences (e.g., service latency, reliability or privacy) render it prohibitive to model the objective function analytically in dynamic IoT environment. In fact, the IoT controller can only have available objective function values at queried points. In this context, the OCO has been extended to the bandit setting by leveraging only point-wise values of objective functions for the gradient estimations, which is referred to as bandit convex optimization (BCO) \cite{BOC1,agarwal2010optimal,shamir2017optimal}. Tailored for partial task offloading strategies among multiple edge servers, BCO with both time-varying costs and constraints was studied in \cite{BOC2}.  On the other hand, aiming at binary computational offloading strategies with such a bandit feedback, multi-armed bandit (MAB) based methods have been popular in MEC systems \cite{MEC_MAB1,MEC_MAB2,MEC_MAB3,sun2017emm}. An online combinatorial bandit upper confidence bound algorithm was proposed in \cite{MEC_MAB1} for the task scheduling to asymptotically minimize the computing delay. The security-aware server selection strategies based on MAB were reported in \cite{MEC_MAB2}. The MAB-based task offloading approach was further adapted to the vehicular edge computing systems in \cite{MEC_MAB3}.

Although achieving promising results, the aforementioned BCO or MAB based works deal only with either continuous or discrete decision variables. In many practical settings though, the analog-amplitude communication and computation resource allocation variables (e.g., transmit power and local computing speed) need to be jointly optimized with discrete variables that capture offloading decisions for optimum MEC performance. Finely discretizing the analog action space (or relaxing the discrete task offloading decisions), renders the existing MAB methods (or the BCO approaches) inaccurate and computationally prohibitive. In addition, the convexity of objective functions commonly assumed in BCO algorithms may not hold in practice \cite{MECsurvey1,MEC2,MEC3,bi2018computation,yan2019optimal}. Although dealing with arbitrary objective functions, MAB methods require to explore every single arm at least once to accumulate sufficient statistics, which may incur sudden performance drops and slow down the learning processes for large MEC networks \cite{MEC_MAB1,MEC_MAB2,MEC_MAB3,sun2017emm}.

Alleviating these limitations, we advocate a novel approach based on Bayesian optimization (BO) \cite{BO_tuto}. BO is a promising methodology for black-box derivative-free (i.e., only function value observations at queried points are available without derivative information) global optimization with well-documented merits, including sample efficiency, uncertainty quantification, and safe exploration \cite{BO_tuto,lu2022surrogate}. The key idea of BO is to build a Bayesian surrogate model (typically, the Gaussian process \cite{Rasmussen2006gaussian, lu2020ensemble, lu2022incremental, polyzos2021ensemble}) for the black-box objective function, guided by which an acquisition function is designed to decide the next function evaluation point. Apart from the applications such as hyperparameter tuning in machine learning \cite{snoek2012practical}, drug discovery \cite{korovina2020chembo}, and robotics \cite{cully2015robots}, BO has been applied to several problems in the context of wireless networks, including radio resource allocation~\cite{BO_RA}, coverage and capacity optimization in cellular networks \cite{dreifuerst2021optimizing}, as well as beam alignment
in mmWave MIMO systems \cite{yang2022bayesian}.
Very recently, targeting video analysis in MEC, a BO-based approach is put forth for edge server and frame resolution selection in \cite{liu2022deep}, where the issue of analog-amplitude communication and computation resource allocation is not accounted for.

Relative to the aforementioned existing works, the present work is the first attempt to develop novel BO-based approaches for the joint optimization of discrete task offloading decisions and analog-amplitude resource allocation strategies in time-varying multi-server multi-user MEC systems with bandit feedback. Specifically, our main contributions are summarized as follows.
\begin{enumerate}
  \item Building on the BO framework for online bandit optimization of categorical and continuous decision variables, a Gaussian process (GP) based surrogate model is adopted for the sought objective function with novel kernel design. The resultant kernel function not only leverages a weighted combination of sum and product compositions of individual kernels over categorical and continuous variables in order to allow for more expressive coupling, but also capitalizes on a temporal kernel to account for unknown dynamics in the black-box function.

  \item With the GP-based surrogate model, an innovative acquisition rule is developed in the time-varying BO scheme to select new optimization variables per iteration. Specifically, given the categorical offloading decisions obtained by the MAB-based method, the analog-amplitude resource allocation variables are determined using the conventional BO-based selection rule.

 % We propose a time-varying BO algorithm by developing a sum and product compositions of the kernels over the discrete and analog-amplitude domains to encode a richer set of their coupling, where a temporal kernel is further comprised to capture the time variation of system dynamics in the Gaussian process (GP) based surrogate model.
 % \item We propose novel BO algorithms in conjunction with the MAB to tackle such challenging online bandit problem with only point-wise objective function values feedback, exacerbated by the combinatorial nature of the task offloading decisions among all WDs and the strong coupling with resource allocation. Specifically, given discrete offloading decisions obtained by the MAB method, the analog-amplitude resource allocation variables will be determined using the BO procedure.
  \item  Under the scenario where each WD reveals its task characterization variables (task computational workload and input data size) at the beginning of each time slot, a generalized contextual time-varying BO scheme is further devised by incorporating the contextual kernel in the GP surrogate model.
 % We further propose a contextual time-varying BO approach by incorporating the contextual kernel in the GP surrogate model, under the scenario where each WD releases its task characterization (i.e., task computational workload and input data size) as the contextual information at the beginning of each time slot.
  \item Numerical simulations under various MEC network sizes demonstrate that our proposed BO approaches benefit from both temporal and contextual information, and exhibit superior performance compared with traditional BO and other representative benchmarks.
\end{enumerate}

The rest of the paper is organized as follows. The system model and problem formulation are presented in Sec. II, following which a novel time-varying BO algorithm for online joint optimization of task offloading and resource allocation under bandit setting is proposed in Sec.~III. Further leveraging observed state information, Sec.~IV develops the contextual time-varying BO approach for dynamic MEC management.
%In Section II, we describe the system model and problem formulation. The time-varying BO algorithm for online joint optimization of task offloading and resource allocation under bandit setting is proposed in Section III. In Section IV, we further develop the contextual time-varying BO approach for dynamic MEC management.
In Sec.~V, the performance of the proposed BO methods is evaluated on synthetic tests. Finally, concluding remarks are made in Sec.~VI.

\emph{Notation:} $(\cdot)^\top$ and $(\cdot)^{-1}$ denote transpose and matrix inverse, respectively, and $\|\mathbf{x}\|$ stands for the $l_2$-norm of a vector $\mathbf{x}$. Besides, $\mathbf{0}_t$, $\mathbf{1}_t$ and $\mathbf{I}_t$ denote the $t\times 1$ all-zero vector, the $t\times 1$ all-one vector and the $t\times t$ identity matrix, respectively. Inequalities for vector $\mathbf{x}>\mathbf{0}$ are entry-wise. $\mathbb{I}(x=x')$ denotes the indicator function taking the value of 1 if $x=x'$, and 0 otherwise. $\mathcal{N}(\mathbf{x};\bm{\mu},\mathbf{K})$ stands for the probability density function (pdf) of a Gaussian random vector $\mathbf{x}$ with mean $\bm{\mu}$ and covariance $\mathbf{K}$.

\section{System Model and Problem Formulation}

\begin{figure}[htb]
\begin{centering}
\includegraphics[scale=0.8]{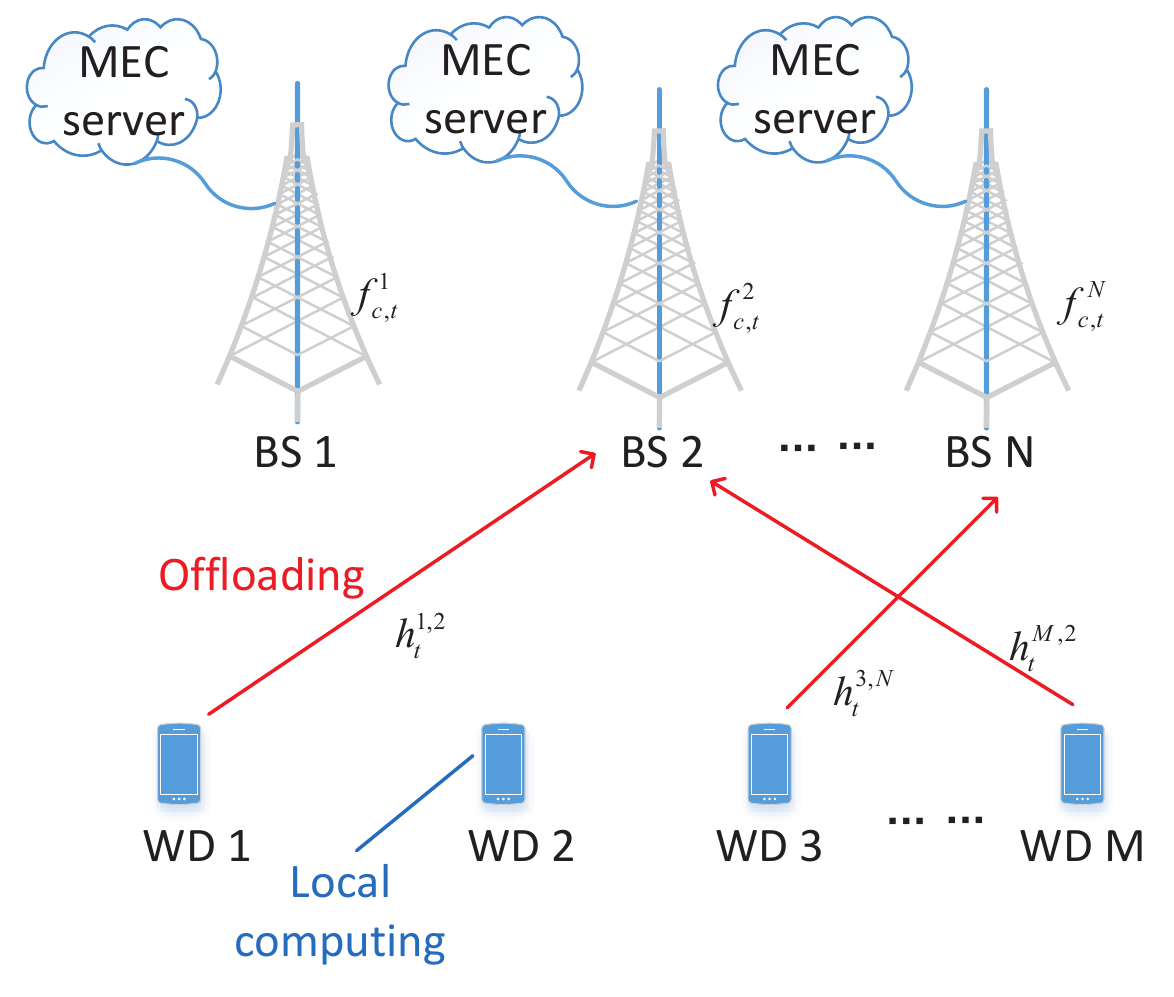}
\vspace{-0.1cm}
 \caption{The considered mobile edge computing (MEC) system with $M$ wireless devices (WDs) and $N$ base stations (BSs).}
\end{centering}
\vspace{-0.1cm}
\end{figure}

Consider a MEC system with $M$ WDs, and $N$ base stations (BSs).
Each BS $n\in\mathcal{N}:=\{1,\ldots,N\}$ is the gateway of edge servers to provide MEC services to the power-limited WDs indexed by $m\in\mathcal{M}:=\{1,\ldots,M\}$.
%Consider that time is discrete and indexed by $t\in\mathcal{T}:=\{1,...,T\}$.
Per slot $t\in\mathcal{T}:=\{1,...,T\}$, the $m$-th WD has a computational task characterized by the pair $(I_{t}^m,L_{t}^m)$, where $I_{t}^m$ denotes the size of input data in bits, and $L_{t}^m$ represents the workload in terms of the total number of CPU cycles to execute the aforementioned task. This WD could either execute its task locally or offload it to one of the BSs, a choice that is henceforth captured by the categorical variable $c_{t}^m\in\{0,1,...,N\}$. Specifically, $c_t^m=0$ indexes local computing, and $c_t^m=n, n\in\mathcal{N}$, stands for offloading task to BS $n$, i.e.,
\begin{align}\label{constraint1}
c^m_t=\left\{
  \begin{array}{ll}
    0, & \hbox{local computing} \\
    n, & \hbox{offloading task to BS $n$}
  \end{array}
\right. \forall m\in\mathcal{M}, n\in\mathcal{N}, t\in\mathcal{T}.
\end{align}
%For the latter, a BS selection variable  $b^{m,n}_t\in\{0,1\}$ will be further introduced with $b^{m,n}_t=1$ representing WD $m$ offloading task to BS $n$, and $b^{m,n}_t=0$ otherwise. To ensure that all the tasks can be executed via either local or edge computing, variables $\{a^m_t\}_{m,t}$ and $\{b^{m,n}_t\}_{m,n,t}$ will conform to the following constraint
%
%\begin{align}\label{constraint1}
% a^{m}_t+ \sum_{n=1}^N b^{m,n}_t =1,~~~a^{m}_t,~b^{m,n}_t\in\{0,1\},~~~ \forall m\in\mathcal{M}, n\in\mathcal{N}, t\in\mathcal{T}.
%\end{align}
%
For both scenarios, the computational overhead per task consists of the execution delay and energy consumption, which will be elaborated as follows.

\subsection{Local Computing}

If WD $m$ chooses to execute its task locally (i.e., $c^m_t=0$) per slot $t$, it has to select the local CPU frequency $f^m_{t}$, based on which the task computing time is given by
\begin{align}
\tau^m_{l,t}= \frac{L^m_t}{f^m_t}
\end{align}
and the corresponding energy consumption is
\begin{align} \label{eq:local_energy}
\epsilon^m_{l,t}=\xi L^m_t(f^m_t)^2
\end{align}
where $\xi$ denotes the effective switched capacitance parameter \cite{MECsurvey1}.

\subsection{Edge Computing}

If WD $m$ alternatively goes for edge computing at BS $n$ per slot $t$, that is, $c^m_t=n$, it must first offload the task using transmit power $p^m_t$. Suppose that the wireless channel coefficient between WD $m$ and BS $n$ for task offloading is $h^{m,n}_t$, and the receiver is corrupted by additive white Gaussian noise (AWGN) with mean zero and variance $\sigma^2$. Here, the wireless channel is assumed to be invariant within each slot and may change
across different slots. Then, the uplink transmission data rate for the sought offloading task is
\begin{align}
R^{m,n}_t=W\log_2(1+\frac{p^m_t |h^{m,n}_t|^2}{\sigma^2})
\end{align}
where $W$ is the identical bandwidth of the dedicated spectral resource block allocated to each WD.  Accordingly, the offloading transmission time is
\begin{align}
\tau^m_{u,t}=\sum_{n=1}^N \frac{\mathbb{I}(c^m_t=n)I^m_t}{R^{m,n}_t}
\end{align}
and the transmission energy consumption of WD $m$ is
\begin{align}
\epsilon^m_{u,t}=p^m_t\tau^m_{u,t}.
\end{align}

For edge computing at BS $n$, the total computation resource per slot $t$ is signified by the CPU frequency $f_{c,t}^n$. Upon receiving all the offloaded tasks, the edge server generates multiple virtual machines (VMs) to  execute the tasks in parallel, and equally partitions $f_{c,t}^n$ to yield $f_{c,t}^{n}/(1+\sum_{m'\in\mathcal{M}/m}\mathbb{I}(c^{m'}_t=n))$ per task.
%Suppose that each edge server generates multiple virtual machines (VMs) for parallel computing, each of which is assigned to execute one of the offloaded tasks. With $f_{c,t}^n$ denoting the given CPU frequency at BS $n$ which also signifies the total computation resource per slot $t$, each VM For edge computing at BS $n$, the given CPU frequency $f_{c,t}^n$ (signifying the total computation resource at BS $n$ per slot $t$) is equally partitioned and shared among the offloaded tasks, yielding $f_{c,t}^{n}/(1+\sum_{m'\in\mathcal{M}/m}\mathbb{I}(c^{m'}_t=n))$ per task.
The edge execution time for WD $m$'s task  is thus
\begin{align}
\tau^m_{c,t}=\sum_{n=1}^N\mathbb{I}(c^m_t=n)\frac{L^m_t (1+\sum_{m'\in\mathcal{M}/m}\mathbb{I}(c^{m'}_t=n))}{f_{c,t}^{n}}.
\end{align}
It is worth mentioning that the time delay for downloading the task output from
the BS to the WD is ignored given the relatively small output data size and strong downlink
transmit power of the BS.

\subsection{Problem Formulation}

Accounting for both local and edge computing, the total time delay for executing the task at WD $m$ per slot $t$ is given by
\begin{align}\label{T}
D^m_t=\mathbb{I}(c^m_t=0)\tau^m_{l,t}+\mathbb{I}(c^m_t\neq 0) (\tau^m_{u,t}+\tau^m_{c,t}).
\end{align}
Here, $D^m_t$ is equal to the local execution time $\tau^m_{l,t}$ if WD $m$ chooses local computing (i.e., $c^m_t=0$). Otherwise, $D^m_t$ in \eqref{T} equals the sum of offloading transmission time $\tau^m_{u,t}$ and the edge computing time $\tau^m_{c,t}$.

Similarly, the energy consumption of WD $m$ per slot $t$ is given by
\begin{align}\label{E}
E^m_t=\mathbb{I}(c^m_t=0) \epsilon^m_{l,t}+\mathbb{I}(c^m_t\neq 0)\epsilon^m_{u,t}
\end{align}
which is $\epsilon^m_{l,t}$ for local computing ($c^m_t=0$) and $\epsilon^m_{u,t}$ otherwise.
%, the total energy cost is the one consumed on local task computing $\epsilon^m_{l,t}$. Otherwise, $E^m_t$ equals the transmission energy consumption for task offloading.

Taking a weighted sum of task execution time delay $D^m_t$ and energy consumption $E^m_t$ yields the energy-delay cost (EDC) per WD $m$ as
\begin{align}\label{EDC}
EDC^m_t(c^m_t,f^m_t,p^m_t)=\beta_{d}D^m_t+\beta_{e}E^m_t
\end{align}
where $\beta_{d},\beta_{e}$ are positive scalars that balance these two costs. For notational brevity, collect the optimization variables in $\mathbf{c}_t:=[c^1_t, \ldots,c^M_t]^\top$,  $\mathbf{p}_t:=[p^1_t,\ldots,p^M_t]^\top$, and $\mathbf{f}_t:=[f^1_t, \ldots, f^M_t]^\top$. The objective is to choose online (at the beginning of each slot $t$) the categorical task offloading decisions (i.e., $\mathbf{c}_t$) and analog-amplitude resource allocation strategies (i.e., $\mathbf{p}_t,\mathbf{f}_t$) minimizing the accumulated EDC across all WDs, that is
\begin{eqnarray}\label{P1}
\mbox{(P1)}~~\min_{\{\mathbf{c}_t,\mathbf{p}_t,\mathbf{f}_t\}_t}&&\sum_{t=1}^T\sum_{m=1}^M EDC^m_t(c^m_t,f^m_t,p^m_t),\nonumber\\
{\rm s.t.}&& c_t^m\in\{0,1,2,...,N\}, ~0<p^m_t\leq P_{peak},~ 0<f^m_t\leq f_{peak},~ \forall m\in\mathcal{M}, t\in\mathcal{T}\nonumber
\end{eqnarray}
where $f_{peak}$ and $P_{peak}$ are the peak local CPU frequency and transmit power of the WDs, respectively.
%For notation brevity, we define categorical variables $\mathbf{c}_t:=[c_t^1,...,c_t^M]^\top$ collecting the offloading decisions $\mathbf{a}_t$ and the edge server selection indicators $\mathbf{b}_t$ of all the WDs at slot $t$, with each $c_t^m\in\{0,1,2,...,N\}$. Specifically, $c_t^m=0$ means local computing of WD $m$, i.e., $a_t^m=1$ and $b^{m,n}_t=0,\forall n$, while $c_t^m=n,n\in\mathcal{N}$, represents task offloading from WD $m$ to the $n$-th BS, i.e., $a_t^m=0$, $b^{m,n}_t=1$, and $b^{m,n'}_t=0, \forall n'\neq n$. Accordingly, it is equivalent to optimize (P1) over the categorical variables $\mathbf{c}_t$ and the analog-amplitude variables $\mathbf{x}_t:=[\mathbf{p}_t^\top, \mathbf{f}_t^\top]^\top$.
By further introducing $\mathbf{x}_t:=[\mathbf{p}_t^\top, \mathbf{f}_t^\top]^\top$ and the reward function $\varphi_t (\mathbf{c}_t,\mathbf{x}_t) :=-\sum_{m=1}^M EDC_t^m$ at slot $t$, (P1) can be equivalently expressed as
\begin{eqnarray}
\mbox{(P2)}~~\max_{\{\mathbf{c}_t,\mathbf{x}_t\}_t}\sum_{t=1}^T\varphi_t (\mathbf{c}_t,\mathbf{x}_t),\nonumber ~~~
{\rm s.t.} ~~\mathbf{c}_t\in\{0,1,2,...,N\}^M,~~\mathbf{0}<\mathbf{x}_t\leq\mathbf{x}_{peak}, \forall t\in\mathcal{T}\nonumber
\end{eqnarray}
where $\mathbf{x}_{peak}:=[P_{peak}\mathbf{1}_M^\top,f_{peak}\mathbf{1}^\top_M]^\top$, and $\mathbf{1}_M$ is the $M$-dimensional all-one column vector.

A major challenge facing (P2) (equivalently (P1)) is that the wireless channels $\{h^{m,n}_t\}$, the edge computing capacities $\{f^n_{c,t}\}$, the computational task characterization $\{I^m_t,L^m_t\}$ are not available; thus, the explicit form of the time-varying EDC function is unknown when making the task offloading and resource allocation decisions $\{\mathbf{c}_t,\mathbf{x}_t\}$ per slot. After performing $\{\mathbf{c}_t,\mathbf{x}_t\}$, only noisy EDC function value (equivalently the realization of $\varphi_t (\mathbf{c}_t,\mathbf{x}_t)$) at that queried point can be acquired at the end of slot $t$. The difficulty of such a bandit setup is further exacerbated by its combinatorial nature that calls for the joint optimization of the categorical $\mathbf{c}_t$ and continuous $\mathbf{x}_t$. To tackle this bandit mix-integer program, novel BO-based approaches will be pursued in the following sections.

\section{Time-Varying BO for Dynamic MEC Management}

BO has well-documented merits in optimizing black-box functions that arise in several settings \cite{BO_tuto}. To account for the temporal variation arising from unknown system dynamics (e.g., changing channel conditions and computing capacities of the edge servers), the slot index $t$ is augmented as an additional input of the sought black-box function, i.e., $\varphi(\mathbf{c}_t,\mathbf{x}_t,t):=\varphi_t(\mathbf{c}_t,\mathbf{x}_t)$.  In short, BO seeks to maximize the black-box $\varphi({\bf z}_t)$ with $\mathbf{z}_t:=[\mathbf{c}_t^\top, \mathbf{x}_t^\top,t]^\top$ by sequentially acquiring function observations using a surrogate model. Collect all the acquired data up to slot $t$ in ${\cal D}_{t}:=\{({\bf z}_\tau, y_\tau)\}_{\tau=1}^{t}$ with $y_\tau$ denoting the possibly noisy observation of $\varphi ({\bf z}_\tau)$. Each BO iteration consists of i) obtaining the function posterior pdf $p(\varphi ({\bf z})|{\cal D}_{t})$ based on the chosen surrogate model using ${\cal D}_{t}$; and, ii) selecting ${\bf z}_{t+1}$ to evaluate at the beginning of slot $t+1$, whose observation $y_{t+1}$ will be acquired at the end of slot $t+1$. In the following, we will introduce the GP-based surrogate model and the acquisition rule for ${\bf z}_{t+1}$, respectively.

\subsection{ GP-based Surrogate Model for Time-Varying Function $\varphi$ and Kernel Design}

As an established Bayesian nonparametric approach, the GP can learn black-box functions with quantifiable uncertainty and sample efficiency, making it suitable for surrogate modeling in BO. Specifically, given data ${\cal D}_{t}$, the goal is to learn the function $\varphi(\cdot)$ that links the input $\mathbf{z}_\tau$ with the scalar output $y_\tau$ as $\mathbf{z}_\tau\rightarrow\varphi(\mathbf{z}_\tau)\rightarrow y_\tau$. Towards this, a GP prior is assumed on the unknown $\varphi$ as $\varphi\thicksim\mathcal{GP}(0,\kappa(\mathbf{z},\mathbf{z}'))$, where $\kappa(\cdot,\cdot)$ is a kernel (covariance) function measuring pairwise similarity of any two inputs. Then, the joint prior pdf of any $t$ function evaluations $\bm{\varphi}_{t}:=[\varphi(\mathbf{z}_1),...,\varphi(\mathbf{z}_{t})]^\top$ at inputs $\mathbf{Z}_{t}:=[\mathbf{z}_1,...,\mathbf{z}_{t}]^\top$ is jointly Gaussian distributed as~\cite{Rasmussen2006gaussian}
\begin{align}\label{GP}
p(\bm{\varphi}_{t}|\mathbf{Z}_{t})=\mathcal{N}(\bm{\varphi}_t;\mathbf{0}_{t},\mathbf{K}_{t}), \forall t
\end{align}
where $\mathbf{K}_t$ is a $t \times t$ covariance matrix with $(\tau,\tau')$-th entry $[\mathbf{K}_t]_{\tau,\tau'}=\mbox{cov}(\varphi(\mathbf{z}_\tau),\varphi(\mathbf{z}_{\tau'})):=\kappa(\mathbf{z}_\tau,\mathbf{z}_{\tau'})$.
The estimation of $\varphi$ relies on the observed outputs $\mathbf{y}_{t}:=[y_1,...,y_{t}]^\top$ that are linked with $\bm{\varphi}_{t}$ through the Gaussian conditional likelihood $p(\mathbf{y}_{t}|\bm{\varphi}_{t},\mathbf{Z}_{t})=\mathcal{N}(\mathbf{y}_{t};\bm{\varphi}_{t},\sigma_o^2\mathbf{I}_{t})$, where $\sigma_o^2$ is the noise variance. Along with the GP prior in \eqref{GP}, one can readily obtain the function posterior pdf $p(\varphi(\mathbf{z})|{\cal D}_t)$ via Bayes' rule as
%we obtain the Gaussian posterior $p(\bm{\varphi}_t|\mathbf{y}_t,\mathbf{Z}_t)\varpropto p(\bm{\varphi}_t|\mathbf{Z}_t)p(\mathbf{y}_t|\bm{\varphi}_t,\mathbf{Z}_t)$ via Bayes' rule.
%For the prediction at a new queried input $\mathbf{z}_{t+1}$ given dataset $\mathcal{D}_t$, we have from \eqref{GP} that $p(\varphi(\mathbf{z}_{t+1})|\bm{\varphi}_t,\mathbf{Z}_t)$ is Gaussian with known mean and covariance \cite{Rasmussen2006gaussian}, based on which the predictive pdf of $\varphi(\mathbf{z}_{t+1})$ is
%
\begin{align}\label{GPpredict}
p(\varphi(\mathbf{z})|{\cal D}_t)= {\cal N}(\varphi(\mathbf{z}); \mu_{t}(\mathbf{z}), \sigma_{t}^2(\mathbf{z}))
%\int p(\varphi(\mathbf{z}_{t+1})|\bm{\varphi}_t,\mathbf{Z}_t)p(\bm{\varphi}_t|\mathbf{y}_t,\mathbf{Z}_t)d\bm{\varphi}_t.
\end{align}
%
%Notice that the posterior $p(\bm{\varphi}_t|\mathbf{y}_t,\mathbf{Z}_t)$ is also Gaussian, yielding the Gaussian predictive pdf $p(\varphi(\mathbf{z}_{t+1})|\mathbf{y}_t,\mathbf{Z}_t)$ in \eqref{GPpredict} with its mean and variance in closed form, i.e.,
%
where its mean and variance have the following closed-form expressions
\begin{align}
\mu_{t}(\mathbf{z})&=\mathbf{k}_{t}^\top(\mathbf{z})(\mathbf{K}_t+\sigma_o^2\mathbf{I}_t)^{-1}\mathbf{y}_t \label{GPmean} \\
\sigma_{t}^2(\mathbf{z})&=\kappa(\mathbf{z},\mathbf{z})-\mathbf{k}_{t}^\top(\mathbf{z})(\mathbf{K}_t+\sigma_o^2\mathbf{I}_t)^{-1}\mathbf{k}_{t}(\mathbf{z}) \label{GPvar}
\end{align}
where $\mathbf{k}_{t}(\mathbf{z}):=[\kappa(\mathbf{z}_{1},\mathbf{z}),...,\kappa(\mathbf{z}_{t},\mathbf{z})]^\top$. Notice that the posterior mean $\mu_t(\mathbf{z})$ is a weighted average of the observed function values $\mathbf{y}_t$, with the weights determined by evaluations of the kernel function at the input values. Besides, the posterior variance $\sigma_{t}^2(\mathbf{z})$ is equal to the prior covariance $\kappa(\mathbf{z},\mathbf{z})$ minus the term corresponding to the variance reduction by observing $\mathbf{y}_t$.

Clearly, the performance of this GP predictor \eqref{GPmean}-\eqref{GPvar} highly hinges on the design of the kernel function $\kappa(\cdot,\cdot)$ over the input space. Accounting for both the continuous $\mathbf{x}_\tau$ for resource allocation and the categorical $\mathbf{c}_\tau$ for task offloading in the function input $\mathbf{z}_\tau$, as well as temporal variations across slots,
%Due to the function input $\mathbf{z}_\tau$ containing both the continuous variables $\mathbf{x}_\tau$ for resource allocation and the categorical variable $\mathbf{c}_\tau$ for task offloading decision, along with the slot index $\tau$ capturing the temporal variation,
three separate kernels are considered, which are  $\kappa_x(\mathbf{x}_\tau,\mathbf{x}_{\tau'})$ over continuous inputs, $\kappa_c(\mathbf{c}_\tau,\mathbf{c}_{\tau'})$ over categorical inputs, and the temporal kernel $\kappa_{temp}(\tau,\tau')$.

Various kernel functions are available for continuous inputs; see~\cite{Rasmussen2006gaussian}.
A popular choice is the class of $Mat\acute{e}rn$ kernels
\begin{align}
\kappa_x^{MT}(\mathbf{x}_{\tau},\mathbf{x}_{\tau'})=\frac{2^{1-\nu}}{\Gamma(\nu)}\left(\frac{\sqrt{2\nu}\|\mathbf{x}_{\tau}-\mathbf{x}_{\tau'}\|}{l}\right)^\nu B_{\nu}\left(\frac{\sqrt{2\nu}\|\mathbf{x}_{\tau}-\mathbf{x}_{\tau'}\|}{l}\right) \label{eq:Matern}
\end{align}
with parameter $\nu >0$ controlling the smoothness of the learning function. The smaller $\nu$ is, the less smooth the sought function is assumed to be. In \eqref{eq:Matern}, $l$ is the characteristic lengthscale, $B_{\nu}$ is a modified Bessel function, and $\Gamma$ is the gamma function. Specifically, as $\nu\rightarrow \infty$, the kernel \eqref{eq:Matern} boils down to the well-known radial
basis function (RBF)
$\kappa^{RBF}_x(\mathbf{x}_\tau,\mathbf{x}_{\tau'}):=\alpha\exp(-\frac{\|\mathbf{x}_\tau-\mathbf{x}_{\tau'}\|^2}{2l^2})$, where the pairwise similarity grows exponentially as a function of the squared distance between any two continuous inputs.

As for categorical variables, we
follow~\cite{cocabo} to adopt the kernel function $\kappa_c(\mathbf{c}_\tau,\mathbf{c}_{\tau'})$ as
%, the pairwise similarity is measured by adopting an indicator function $\mathbb{I}(\cdot,\cdot)$ \cite{cocabo}, i.e.,
%
\begin{align}\label{cat}
\kappa_c(\mathbf{c}_\tau,\mathbf{c}_{\tau'})=\frac{\omega}{M}\sum_{m=1}^M\mathbb{I}(c_\tau^m=c_{\tau'}^m)
\end{align}
where
%the indicator function $\mathbb{I}(c_\tau^m=c_{\tau'}^m)$ takes the value of $1$ if $c_\tau^m=c_{\tau'}^m$, and $0$ otherwise.
$\omega$ is the categorical kernel variance. Note that the categorical kernel defined in \eqref{cat} is a special case of the RBF kernel with $\alpha=1$ and $l\rightarrow 0$. To allow for a richer set of couplings between the continuous and categorical domains, a mixture of the sum and product compositions of the two kernels $\kappa_x$ and $\kappa_c$ is proposed for the kernel function $\kappa_{x,c}$ over continuous and categorical variables \cite{cocabo}, i.e.,
\begin{align}\label{cx}
\kappa_{x,c}([\mathbf{x}_\tau^\top,\mathbf{c}_\tau^\top]^\top, [\mathbf{x}_{\tau'}^\top, \mathbf{c}_{\tau'}^\top]^\top)=(1-\lambda)[\kappa_c(\mathbf{c}_\tau,\mathbf{c}_{\tau'})+\kappa_x(\mathbf{x}_\tau,\mathbf{x}_{\tau'})]+\lambda\kappa_c(\mathbf{c}_\tau,\mathbf{c}_{\tau'})\kappa_x(\mathbf{x}_\tau,\mathbf{x}_{\tau'})
\end{align}
where $\lambda\in[0,1]$ weighs the contributions from the sum and product compositions of $\kappa_c$ and $\kappa_x$.
When $\lambda=0$, only the sum composition exists in~\eqref{cx}, leading to independence of the black-box function $\varphi$ over the continuous and categorical domains with limited expressiveness. On the other hand, the pure product composition with $\lambda=1$ will take the value of $0$ if there is no pairwise overlap between two categorical variables $\mathbf{c}_\tau$ and $\mathbf{c}_{\tau'}$, that is, $\kappa_c (\mathbf{c}_\tau, \mathbf{c}_\tau')=0$ according to \eqref{cat}, thus preventing the GP model from learning. Towards overcoming the aforementioned two limitations, one can leverage a weighted combination of the sum and product components with $0<\lambda<1$ in~\eqref{cx}.

%\changeQ{Note that onefold additive kernel combination (i.e., $\lambda=0$) indicates independence of the black-box function $\varphi$ over the continuous and categorical domains, leading to limited correlation expressiveness. Additionally, with no pairwise overlap between two categorical variables $\mathbf{c}_\tau$ and $\mathbf{c}_{\tau'}$, $\kappa_c=0$ according to \eqref{cat}, which prevents pure product kernel combination with  $\lambda=1$ from encoding covariance over such hybrid input space. Leveraging the strength of both additive and product kernel combination via \eqref{cx} overcomes the above limitations.}

To further capture the temporal variation of the black-box function $\varphi$ due to the unknown system dynamics, the following temporal kernel function $\kappa_{temp}(\tau,\tau')$ is adopted based on~\cite{bogunovic2016time}
\begin{align}\label{temp}
\kappa_{temp}(\tau,\tau')=(1-\rho)^{\frac{|\tau-\tau'|}{2}}
\end{align}
where $\rho\in[0,1]$ is the hyperparameter that controls the level of temporal dynamics in the learning function $\varphi$. The larger the value of $\rho$, the more frequently $\varphi$ varies over time. In particular, when $\rho=0$, $\kappa_{temp}(\tau,\tau')=1$ for any $(\tau, \tau')$, thus inducing no dynamics in $\varphi$.

Henceforth, applying the product composition of $\kappa_{x,c}$ \eqref{cx} and $\kappa_{temp}$ \eqref{temp} yields the overall kernel function given by
%Accordingly, the overall kernel $\kappa(\mathbf{z}_\tau,\mathbf{z}_{\tau'})$ over the input space is proposed as the product combination of $\kappa_{x,c}$ in \eqref{cx} and $\kappa_{temp}$ in \eqref{temp},
%
\begin{align}\label{eq:k_z}
\kappa(\mathbf{z}_\tau,\mathbf{z}_{\tau'})=\kappa_{temp}(\tau,\tau')\kappa_{x,c}([\mathbf{x}_\tau^\top,\mathbf{c}_\tau^\top]^\top, [\mathbf{x}_{\tau'}^\top, \mathbf{c}_{\tau'}^\top]^\top).
\end{align}
It can be observed that the temporal kernel imposes a scaling factor on $\kappa_{x,c}$ based on the time separation of any pair of inputs.
%The temporal kernel can be viewed as a weight on $\kappa_{x,c}$ indicating the freshness of the samples in the dataset.
This agrees well with intuition that inputs that are well separated in time (i.e., large $|\tau-\tau'|$) yield less correlated function values for $\rho\neq 0$.
%where a older sample with larger $|\tau-\tau'|$ is assigned a smaller weight and contributes less to the final values of $\mu_{t}$ and $\sigma_{t}^2$.
%Besides, $\rho$ in the temporal kernel is a parameter controlling time variation of black-box function $\varphi$, where a larger $\rho$ represents a black-box function with larger time variation.  In particular, $\rho=0$ corresponds to the traditional time-invariant BO.

\noindent
\textbf{Remark 1 (Learning the GP hyperparameters).} The GP hyperparameters, collected in $\bm{\theta}$ that consists of
the characteristic length-scale $l$, categorical kernel variance $\omega$, and the noise variance $\sigma_o^2$, are optimized by maximizing the log marginal likelihood~\cite{Rasmussen2006gaussian}
\begin{align}
\mathcal{L}(\bm{\theta})&:=\log p(\mathbf{y}_t|\mathbf{Z}_t)=\log\left(\int p(\mathbf{y}_{t}|\bm{\varphi}_{t},\mathbf{Z}_{t})p(\bm{\varphi}_{t}|\mathbf{Z}_{t})d\bm{\varphi}_{t}\right)\nonumber\\
&=-\frac{1}{2}\mathbf{y}_{t}^\top(\mathbf{K}_t+\sigma_o^2\mathbf{I}_t)^{-1}\mathbf{y}_t-\frac{1}{2}\log|\mathbf{K}_t+\sigma_o^2\mathbf{I}_t|-\frac{t}{2}\log2\pi
\label{LML}
\end{align}
where the first term involving the observations represents the data-fit; the second term indicates the complexity penalty; and, the last term is a normalization constant. Accordingly, the gradient of the $\mathcal{L}(\bm{\theta})$ with respect to the hyperparameters $\bm{\theta}$ is given by
\begin{align}\label{LML_grad}
\frac{\partial\mathcal{L}(\bm{\theta})}{\partial\bm{\theta}}=\frac{1}{2}\mathbf{y}_{t}^\top(\mathbf{K}_t+\sigma_o^2\mathbf{I}_t)^{-1}\frac{\partial(\mathbf{K}_t+\sigma_o^2\mathbf{I}_t)}{\partial\bm{\theta}}(\mathbf{K}_t+\sigma_o^2\mathbf{I}_t)^{-1}\mathbf{y}_t-\frac{1}{2}\mbox{tr}\left((\mathbf{K}_t+\sigma_o^2\mathbf{I}_t)^{-1}\frac{\partial(\mathbf{K}_t+\sigma_o^2\mathbf{I}_t)}{\partial\bm{\theta}}\right)
\end{align}
based on which the gradient-based optimizer is adopted to learn $\bm{\theta}$ every $\delta$ time slots.

\subsection{Acquisition for $\mathbf{z}_{t+1}$ Based on GP Surrogate Model}

Having available GP-based posterior function model~\eqref{GPpredict} with the form of kernel function specified by~\eqref{eq:k_z} at slot $t$, one is ready to select the next decisions $\mathbf{z}_{t+1}$. Coping with both categorical and continuous variables, this is certainly a nontrivial task, but can fortunately be handled by relying on the MAB framework. Since the cardinality of the categorical variables is exponential with respect to the number $M$ of WDs, a scalable multi-agent MAB approach will be leveraged with each WD $m$ acting as an agent simultaneously and independently determining its local task offloading decision $c_{t}^m\in\{0,1,...,N\}$. As the overall reward function in the resultant MAB framework does not follow any statistical distribution, it is more sensible to rely on the adversarial MAB framework and adopt as the action selection rule the well-known exponential-weight algorithm for exploration and exploitation (EXP3) \cite{EXP3}. Per slot $t$, EXP3 maintains an unnormalized weight vector $\mathbf{w}_{t}^m:=[w_{t}^m(0),w_{t}^m(1),...,w_{t}^m(N)]^\top$ for each WD $m$ to guide the selection of its action. Next, we will delineate how each acquisition step of the time-varying BO selects categorical ${\bf c}_{t+1}$ and continuous ${\bf x}_{t+1}$ with the help of EXP3.

%exponentially increasing action space $(N+1)^M$ with respect to the number of WDs $M$, a scalable multi-agent MAB approach is considered with each WD $m$ acting as an agent simultaneously determining its local task offloading decision $c_{t}^m\in\{0,1,...,N\}$

%Specifically, with both categorical and analog variables, this selection proceeds in two steps. Relying on a multi-agent MAB approach with each WD viewed as an agent, the discrete task offloading decisions ${\bf c}_{t}$ are first determined using the exponential-weight algorithm for exploration and exploitation (EXP3) \cite{EXP3}. Subsequently, the analog-amplitude resource allocation variables ${\bf x}_{t}$ are  obtained using the celebrated upper confidence bound (UCB) based acquisition rule~\cite{srinivas2012information}. The details of these two steps are delineated in the ensuing subsections.

%The algorithm we obtain discrete task offloading decisions ${\bf c}_{t}$ given analog-amplitude variables ${\bf x}_{t}$ builds on the EXP3 due to its applicability to general adversarially-defined function $\varphi$ without identically and independently distributed (i.i.d.) function assumption as in UCB and $\varepsilon$-greedy MAB algorithms. Due to the exponentially increasing action space $(N+1)^M$ with respect to the number of WDs $M$, a scalable multi-agent MAB approach is considered with each WD $m$ acting as an agent simultaneously determining its local task offloading decision $c_{t}^m\in\{0,1,...,N\}$ through EXP3.

\subsubsection{Acquisition for Categorical Task Offloading Decisions}

Given ${\bf w}_{t}^m$ from the end of slot $t$, each agent $m$ in EXP3 draws its action $c_{t+1}^m$ randomly according to the probability vector $\mathbf{q}_{t}^m:=[q_{t}^m(0),q_{t}^m(1),...,q_{t}^m(N)]^\top$ with\cite{EXP3}
\begin{align}
q_{t}^m(k)=(1-\gamma)\frac{w_{t}^m(k)}{\sum_{k'=0}^Nw_{t}^m(k')}+\frac{\gamma}{N+1}, \forall k\in\{0,1,...,N\} \label{mabprob}
\end{align}
where $\gamma\in(0,1]$ is the coefficient that balances {\it exploitation} given by the normalized weight in the first factor and {\it exploration} from the uniform probability in the second term. Specifically, by including the uniform distribution, EXP3 allows all $N+1$ decisions to be explored per agent (WD) so as to get good reward estimates.

\subsubsection{Acquisition for Analog-Amplitude Resource Allocation Decisions}

With the categorical task offloading decisions $\mathbf{c}_{t+1}$ at hand, the analog-amplitude resource allocation decisions $\mathbf{x}_{t+1}$ are selected by finding the maximizer of the celebrated upper confidence bound (UCB)-based acquisition function as~\cite{srinivas2012information} %Recall that given $\mathcal{D}_{t-1}$, the GP-based surrogate model provides a Gaussian posterior pdf for the black-box function $\varphi$ with mean $\mu_{t-1}$ in \eqref{GPmean} and variance $\sigma^2_{t-1}$ in \eqref{GPvar}. The UCB-based acquisition rule is to maximize the following linear combination of the posterior mean and variance as
\begin{align}\label{acq_problem}
\mathbf{x}_{t+1} = \underset{0<\mathbf{x}\leq\mathbf{x}_{peak}}{\arg\max} \ u_{t+1}(\mathbf{x}|\mathcal{D}_{t},\mathbf{c}_{t+1},t+1):=\mu_{t}(\mathbf{x},\mathbf{c}_{t+1},t+1)+\sqrt{\zeta_{t+1}}\sigma^2_{t}(\mathbf{x},\mathbf{c}_{t+1},t+1)
\end{align}
where the coefficient $\zeta_{t+1}\geq 0$ nicely balances the exploitation and exploration that are signified by the posterior mean  $\mu_{t}$~\eqref{GPmean} and variance $\sigma^2_{t}$~\eqref{GPvar}, respectively.
%Specifically, the posterior mean is maximized to exploit the past experience in $\mathcal{D}_{t-1}$, while the exploration is achieved by maximizing the quantified uncertainty $\sigma^2_{t-1}$.
With closed-form expressions of $\mu_{t}$ and $\sigma^2_{t}$ at hand, one can readily solve~\eqref{acq_problem} via off-the-shelf gradient-based solvers.

\subsubsection{Weight Update in EXP3}
%Having acquired $({\bf c}_t, {\bf x}_t)$, the central controller will implement them in the MEC system to obtain the observed reward $y_t$. Then, EXP3 will update the weight as
Upon deploying $({\bf c}_{t+1}, {\bf x}_{t+1})$ into the MEC system to yield the observed reward $y_{t+1}$,  EXP3 capitalizes on the importance sampling rule to obtain an unbiased estimate of the reward value as
\begin{align}\label{mabreward}
\hat{\varphi}_{t+1}^m(k)=\frac{y_{t+1}\mathbb{I}(c_{t+1}^m=k)}{q_{t}^m(k)}, \forall k\in\{0,1,...,N\}, m\in\mathcal{M}
\end{align}
based on which the corresponding weight is updated using the exponential rule as
\begin{align}
w_{t+1}^m(k)&=w_{t}^m(k)\exp\left(\frac{\gamma\hat{\varphi}_{t+1}^m(k)}{N+1}\right)\nonumber\\
&=w_{0}^m(k)\exp\left(\frac{\gamma\sum_{\tau=1}^{t+1}\hat{\varphi}_{\tau}^m(k)}{N+1}\right), \forall k\in\{0,1,...,N\}, m\in\mathcal{M}\label{mabweight}.
\end{align}
It is evident that $w_{t+1}^m(k)$ summarizes the cumulative rewards up to slot $t+1$ for action $k$ under WD $m$, and thus represents the effect of exploitation in~\eqref{mabprob}.

%After the edge computing tasks are completed at slot $t$, only the joint reward function $\varphi_{t}$ depending on the selected actions ${\bf c}_{t}$ of all agents is observed, which further coordinates the EXP3 update of each agent. To complement such partial feedback of the reward, EXP3 utilizes an importance sampling type of reward estimates given by
%\begin{align}
%w_{t}^m(k)&=w_{t-1}^m(k)\exp\left(\frac{\gamma\hat{\varphi}_{t}^m(k)}{N+1}\right)\nonumber\\
%&=\exp\left(\frac{\gamma\sum_{\tau=1}^t\hat{\varphi}_{\tau}^m(k)}{N+1}\right), \forall k\in\{0,1,...,N\}.\label{mabweight}
%\end{align}
%The indicator function in the numerator represents that only one of $(N+1)$ entries of the estimated reward vector at each agent is nonzero, while the probability $q_{t}^m(k)$ in the denominator ensures unbiasedness and compensates the rewards of the actions that are unlikely to be picked. We rely on \eqref{mabreward} to update the next action distribution $\mathbf{q}_{t+1}^m$ of agent $m$ given by \cite{EXP3}
The pseudo-code of the overall time-varying BO approach is summarized in Algorithm 1.

\begin{algorithm}[htb]
\caption{Time-Varying BO for Dynamic MEC Management}
\begin{algorithmic}[1]
\STATE \textbf{Initialization:} $\mathcal{D}_0$, $w_0^m(k)=1, \forall k\in\{0,1,...,N\}, m\in\mathcal{M}$;
\FOR{$t=0:T-1$}
\IF{$t\mod\delta=1$}
\STATE Learn GP hyperparameters $\bm{\theta}$ via multi-started gradient descent using \eqref{LML_grad};
\ENDIF
\STATE Calculate the posterior mean $\mu_{t}$ and variance $\sigma^2_{t}$ according to \eqref{GPmean}--\eqref{GPvar} given $\mathcal{D}_{t}$;
\STATE Compute the action distribution $\mathbf{q}_{t}^m,\forall m\in\mathcal{M}$ according to \eqref{mabprob};
\STATE Draw the discrete task offloading decision $c_{t+1}^m$ randomly according to $\mathbf{q}_t^m, \forall m\in\mathcal{M}$;
\STATE Acquire the analog-amplitude resource allocation decisions $\mathbf{x}_{t+1}$ by solving \eqref{acq_problem};
\STATE Deploy decisions $\mathbf{z}_{t+1}:=[\mathbf{c}_{t+1}^\top,\mathbf{x}_{t+1}^\top,t+1]^\top$ to MEC system to observe $y_{t+1}$;
\STATE $\mathcal{D}_{t+1}=\mathcal{D}_{t}\cup\{(\mathbf{z}_{t+1},y_{t+1})\}$ and update  $w_{t+1}^m(k)$ via \eqref{mabweight}, $\forall k\in\{0,1,...,N\}, m\in\mathcal{M}$;
\ENDFOR
\end{algorithmic}
\end{algorithm}

%These two steps offer quantifiable uncertainty and sample efficiency, thanks to the Gaussian process (GP) that is adopted to model $\varphi ({\bf z})$ a priori, that is $\varphi\thicksim\mathcal{GP}(0,\kappa(\mathbf{z},\mathbf{z}'))$, where the kernel function $\kappa(\mathbf{z},\mathbf{z}')$ is selected as $\kappa(\mathbf{z},\mathbf{z}')=(1-\lambda)[\kappa_c(\mathbf{c},\mathbf{c}')+\kappa_x(\mathbf{x},\mathbf{x}')]+\lambda\kappa_c(\mathbf{c},\mathbf{c}')\kappa_x(\mathbf{x},\mathbf{x}')$, while $\lambda\in[0,1]$ weighs the sum and product compositions of $\kappa_c$ and $\kappa_x$, the kernel functions over discrete and analog variables. Further connecting $\varphi ({\bf z}_\tau)$ with $y_\tau$ through the Gaussian likelihood, one can readily obtain $p(\varphi ({\bf z})|{\cal D}_t)$ analytically~\cite{Rasmussen2006gaussian}, based on which ${\bf z}_{t+1}$ can be selected. With both discrete and analog variables, this selection proceeds in two steps. Relying on a multi-agent MAB approach with each WD viewed as an agent, the discrete variables ${\bf c}_{t+1}$ are first determined using the EXP3 scheme \cite{EXP3}. Subsequently, the analog-amplitude variables ${\bf x}_{t+1}$ are  obtained using the celebrated upper confidence bound (UCB) based acquisition rule~\cite{srinivas2012information}. The details of the proposed BO approach will be presented in the full version of the paper.

\section{Contextual Time-Varying BO for Dynamic MEC Management}

So far, we have introduced a time-varying BO approach for dynamic MEC management under the bandit setting, where the temporal dynamics of the black-box reward function $\varphi$ is captured by incorporating the temporal kernel. In some scenarios, in addition to the observed reward value, one could have access to a subset of the system state. Here,  each WD $m$ could report its task characterization variables $(I_t^m, L_t^m)$ to the central controller per slot $t$.
%Towards leveraging such observed state information, this section aims to generalize the previous time-varying BO approach
The goal of this section is then to generalize the time-varying BO approach for more informed decision-making by leveraging such state information, which will also be termed as ``context" hereafter.

%In Section III, we propose a time-varying Bayesian optimization approach for dynamic MEC management problem under bandit setting, where the time variation of the black-box reward function $\varphi$ is captured by incorporating the temporal information. This section deals with the scenario where each WD reports its task characterization $(I_t^m,L_t^m)$ as the contextual information at the beginning of each slot when determining task offloading and resource allocation decisions $\mathbf{z}_t$, based on which a contextual time-varying Bayesian optimization algorithm is proposed. \changeQ{It is redundant, try to compress and summarize}

In the resultant contextual time-varying BO approach, the black-box reward function $\varphi(\bar{\mathbf{z}}_t)$ has the augmented input $\bar{\mathbf{z}}_t:=[\mathbf{z}_t^\top, \mathbf{s}_t^\top]^\top$, where $\mathbf{s}_t$ is the context vector that collects the observed state information as $\mathbf{s}_t:=[I_t^1,...,I_t^M,L_t^1,...,L_t^M]^\top$. As with the time-varying BO approach in the previous section, the generalized counterpart here still consists of two steps per iteration, namely, GP-based surrogate model learning and the acquisition of new decisions.

For the former, a GP prior is postulated for $\varphi$ as $\varphi\sim\mathcal{GP}(0,\bar{\kappa}(\bar{\mathbf{z}},\bar{\mathbf{z}}'))$, where
the kernel function $\bar{\kappa}$ has to be adapted to capture correlation from the contextual input. Inspired by~\cite{krause2011contextual}, $\bar{\kappa}(\bar{\mathbf{z}}_\tau,\bar{\mathbf{z}}_{\tau'})$ is proposed as the product combination of three separate kernels given by
\begin{align}
\bar{\kappa}(\bar{\mathbf{z}}_\tau,\bar{\mathbf{z}}_{\tau'})=\kappa_s(\mathbf{s}_\tau,\mathbf{s}_{\tau'})\kappa_{temp}(\tau,\tau')\kappa_{x,c}([\mathbf{x}_\tau^\top,\mathbf{c}_\tau^\top]^\top, [\mathbf{x}_{\tau'}^\top, \mathbf{c}_{\tau'}^\top]^\top)
\end{align}
where $\kappa_s(\mathbf{s}_\tau,\mathbf{s}_{\tau'})$ is the contextual kernel over the observed context variables, and  $\kappa_{x,c}$ and $\kappa_{temp}$ are given by \eqref{cx} and \eqref{temp}. Given the GP prior and a set of input-output data pairs $\bar{\mathcal{D}}_{t}:=\{(\bar{\mathbf{z}}_\tau,y_\tau)\}_{\tau=1}^{t}$, the posterior pdf for the reward function is given by (cf.~\eqref{GPpredict})
\begin{align}
p(\varphi(\bar{\mathbf{z}})|\bar{\mathcal{D}}_{t}) =  {\cal N}(\varphi(\bar{\mathbf{z}}); \bar{\mu}_{t}(\bar{\mathbf{z}}), \bar{\sigma}_{t}^2(\bar{\mathbf{z}}))\label{eq: GP_post_context}
\end{align}
where the closed-form expressions of the mean $\bar{\mu}_t$ and variance $\bar{\sigma}_{t}^2$ can be obtained similarly as in~\eqref{GPmean}--\eqref{GPvar} by including the context vectors in the input, i.e.,
\begin{align}
\bar{\mu}_{t}(\bar{\mathbf{z}})&=\bar{\mathbf{k}}_{t}^\top(\bar{\mathbf{z}})(\bar{\mathbf{K}}_t+\sigma_o^2\mathbf{I}_t)^{-1}\mathbf{y}_t \label{GPmean_contex} \\
\bar{\sigma}_{t}^2(\bar{\mathbf{z}})&=\bar{\kappa}(\bar{\mathbf{z}},\bar{\mathbf{z}})-\bar{\mathbf{k}}_{t}^\top(\bar{\mathbf{z}})(\bar{\mathbf{K}}_t+\sigma_o^2\mathbf{I}_t)^{-1}\bar{\mathbf{k}}_{t}(\bar{\mathbf{z}}). \label{GPvar_contex}
\end{align}
Here $\bar{\mathbf{k}}_{t}(\bar{\mathbf{z}}):=[\bar{\kappa}(\bar{\mathbf{z}}_{1},\bar{\mathbf{z}}),...,\bar{\kappa}(\bar{\mathbf{z}}_{t},\bar{\mathbf{z}})]^\top$ and $\bar{\mathbf{K}}_t$ is the $t\times t$ covariance matrix with $(\tau,\tau')$-th entry $[\bar{\mathbf{K}}_t]_{\tau,\tau'}:=\bar{\kappa}(\bar{\mathbf{z}}_\tau,\bar{\mathbf{z}}_{\tau'})$. Similar as Remark 1 in Sec.~III-A, the GP hyperparameters $\bar{\bm{\theta}}$ is optimized every $\delta$ slots via log marginal likelihood maximization using \eqref{LML_grad}.

As for the acquisition of task offloading and resource allocation decisions for slot $t+1$, contextual time-varying BO proceeds as in Sec.~III-B by first selecting the categorical ${\bf c}_{t+1}$ via the EXP3 approach based on the multi-agent MAB framework, and then choosing the continuous $\mathbf{x}_{t+1}$ using the UCB rule. Here, the latter has to take into account the observed context vector $\mathbf{s}_{t+1}$, thus yielding $\mathbf{x}_{t+1}$ given by
%Similarly, the discrete task offloading decisions $\mathbf{c}_t$ are first selected via the multi-agent EXP3 MAB approach proposed in Section III B1, followed by which the analog-amplitude resource allocation strategy $\mathbf{x}_t$ is obtained using the UCB-based acquisition rule by incorporating the contextual information $\mathbf{s}_t$. Specifically, by directly utilizing the closed-form posterior mean and variance, the UCB-based acquisition strategy aims at solving the following optimization problem
%
\begin{align}\label{acq_problem_context}
\mathbf{x}_{t+1} = \underset{0<\mathbf{x}\leq\mathbf{x}_{peak}}{\arg\max} \ \bar{u}_{t+1}(\mathbf{x}|\bar{\mathcal{D}}_{t},\mathbf{c}_{t+1},\mathbf{s}_{t+1},t\!+\!1):=\bar{\mu}_{t}(\mathbf{x},\mathbf{c}_{t+1},\mathbf{s}_{t+1},t\!+\!1)+\sqrt{\bar{\zeta}_{t+1}}\bar{\sigma}^2_{t}(\mathbf{x},\mathbf{c}_{t+1},\mathbf{s}_{t+1},t\!+\!1)
\end{align}
where $\bar{\zeta}_{t+1}\geq 0$ is the coefficient that balances exploration and exploitation.
Please refer to Algorithm 2 for the detailed implementation of the contextual time-varying BO approach.

\begin{algorithm}[htb]
\caption{Contextual Time-Varying BO for Dynamic MEC Management}
\begin{algorithmic}[1]
\STATE \textbf{Initialization:} observation dataset $\bar{\mathcal{D}}_0$ and $w_0^m(k)=1, \forall k\in\{0,1,...,N\}, m\in\mathcal{M}$;
\FOR{$t=0:T-1$}
\IF{$t\mod\delta=1$}
\STATE Learn GP Hyperparameters $\bar{\bm{\theta}}$ via multi-started gradient descent using \eqref{LML_grad};
\ENDIF
\STATE Calculate the mean $\bar{\mu}_{t}$ and variance $\bar{\sigma}^2_{t}$ in the posterior pdf~\eqref{eq: GP_post_context} according to \eqref{GPmean_contex}--\eqref{GPvar_contex};
\STATE Observe the contextual information $\mathbf{s}_{t+1}$;
\STATE Compute the action distribution $\mathbf{q}_t^m, \forall m\in\mathcal{M}$ according to \eqref{mabprob};
\STATE Draw the discrete task offloading decision $c_{t+1}^m$ randomly according to $\mathbf{q}_t^m, \forall m\in\mathcal{M}$;
\STATE Acquire the analog-amplitude resource allocation decisions $\mathbf{x}_{t+1}$ by solving \eqref{acq_problem_context};
\STATE Deploy decisions $(\mathbf{c}_{t+1},\mathbf{x}_{t+1})$ to the MEC system to observe $y_{t+1}$;
\STATE $\bar{\mathcal{D}}_{t+1}=\bar{\mathcal{D}}_{t}\cup\{(\bar{\mathbf{z}}_{t+1},y_{t+1})\}$ and update  $w_{t+1}^m(k)$  via~\eqref{mabweight}, $\forall k\in\{0,1,...,N\}, m\in\mathcal{M}$;
\ENDFOR
\end{algorithmic}
\end{algorithm}

\section{Simulation Results}

In this section, numerical tests were conducted to evaluate the performance of the proposed BO approaches for dynamic MEC management. In the multi-user multi-server MEC system with $M$ WDs and $N$ BSs, the time-varying wireless channel $h_t^{m,n}$ from WD $m$ to BS $n$ is modelled as Rician fading channel
\begin{align}
h_t^{m,n}=\sqrt{\frac{K}{K+1}}h_{t,LoS}^{m,n}+\sqrt{\frac{1}{K+1}}h_{t,NLoS}^{m,n}, \forall m,n,t \label{eq: h_t}
\end{align}
where $h_{t,LoS}^{m,n}$ denotes the deterministic line of sight (LoS) component determined by the locations of BS $n$ and WD $m$; $h_{t,NLoS}^{m,n}$ stands for the non-LoS component following the independent and identically
distributed (i.i.d.) standard Gaussian distribution; and $K\geq 0$ is the Rician factor representing the ratio of the power in the LoS component to the power in the non-LoS component. Note that a larger $K$ implies milder fading effect. The total  average channel gain follows the free-space path loss model $|\bar{h}_t^{m,n}|^2=A_d(\frac{3\times 10^8}{4\pi \phi d_{m,n}})^{PL}, \forall t$, where $A_d=4.11$ denotes the antenna gain, $\phi=915$ MHz is the carrier frequency, $d_{m,n}$ represents the distance (measured by meters) between WD $m$ and BS $n$, and $PL=3$ signifies the pass loss exponent.

In addition,  the means of time-varying edge CPU frequencies $\{f_{c,t}^n\}_{n,t}$, task computational workloads $\{L_{t}^m\}_{m,t}$, and task input data sizes $\{I_{t}^m\}_{m,t}$ are 26 GHz, 125 Mcycles, and 1250 KBytes, respectively \cite{MEC2,MEC1,bi2018computation,yan2019optimal}. Specifically, the generation rules are as follows
\begin{subequations}
\begin{align}
 f_{c,t}^n &= (\bar{f}_{c}+\tilde{f}_{c,t}^n)\times 10^9 ~\mbox{Hz} \  \  \forall n,t \\
 L_{t}^m &= (\bar{L} +\tilde{L}_{t}^m)\times 10^6 ~\mbox{Cycles} \ \ \forall m,t \ \\
I_{t}^m &= (\bar{I}+\tilde{I}_{t}^m)\times 10^4 ~\mbox{Bytes}\ \ \  \forall m,t
\end{align}
\end{subequations}
where $\bar{f}_{c} = 26$, $\bar{L} = 125$, $\bar{I} = 125$, and the dynamic components $\tilde{f}_{c,t}^n$, $\tilde{L}_{t}^m$, and $\tilde{I}_{t}^m$  are evolved based on the following first-order Markovian processes
%\changeQ{Comment on notation: $\bar{}$ and $\hat{}$ usually refer to average and estimate of some random variable. It is not suggested to use them to represent random variables.}
%
\begin{subequations}\label{eq: state_gen}
\begin{align}
\tilde{f}_{c,1}^n&={e}_{f, 1}^n \ \  \tilde{L}_{1}^m = {e}_{L,1}^m\ \  \tilde{I}_{1}^m= {e}_{I, 1}^m \\
\tilde{f}_{c,t+1}^n &=\sqrt{1-\eta}\tilde{f}_{c,t}^n+\sqrt{\eta}{e}_{f,t+1}^n, \ {e}_{f,t+1}^n\sim \mathcal{N}(0,3)  \\
\tilde{L}_{t+1}^m &=\sqrt{1-\eta}\tilde{L}_{t}^m+\sqrt{\eta}{e}_{L,t+1}^m,  \ {e}_{L,t+1}^m\sim\mathcal{N}(0,3) \\
\tilde{I}_{t+1}^m &=\sqrt{1-\eta}\tilde{I}_{t}^m+\sqrt{\eta}{e}_{I, t+1}^m,  \ {e}_{I, t+1}^m \sim\mathcal{N}(0,3)
\end{align}
\end{subequations}
where the process noises ${e}_{f,t}^n$, ${e}_{L,t}^m$, and ${e}_{I,t}^m$ are i.i.d., and $\eta\in[0,1]$ is the parameter adjusting the level of temporal dynamics in these system state variables. In particular, $\eta=0$ represents the time-invariant scenario, while $\eta=1$ indicates the independent system dynamics across time slots \cite{bogunovic2016time}. %Note that the units of $\{f_{c,t}^n\}_{n,t}$, $\{L_{t}^m\}_{m,t}$, and $\{I_{t}^m\}_{m,t}$ are respectively GHz, Mcycles, and 10KBytes.

% $\{\hat{f}_{c,t}^n\}_{n,t}$, $\{\hat{L}_{t}^m\}_{m,t}$, $\{\hat{I}_{t}^m\}_{m,t}$ be independent random variables with $\hat{f}_{c,t}^n, \hat{L}_{t}^m, \hat{I}_{t}^m\sim\mathcal{N}(0,3)$, we model

%\changeQ{comment: cite the eq. number of the parameters.}
Besides, the peak transmit power $P_{peak}$ and computational frequency $f_{peak}$ of each WD are equal to $100$ mW and $10^8$ Hz, respectively. To be aligned with commercial practise, the computing efficiency coefficient
%We consider the commercial WDs in practice with the computing efficiency coefficient
$\xi$ of the WDs in \eqref{eq:local_energy} is chosen as $\xi=10^{-26}$ \cite{miettinen2010energy}. We set the channel additive white Gaussian noise power $\sigma^2=10^{-10}$ W, and the bandwidth $W=2$ MHz. The prior weights of the time delay and energy consumption cost of the WDs in \eqref{EDC} are set as $\beta_d=\beta_e=0.5$.

For the proposed (contextual) time-varying BO approaches, the $Mat\acute{e}rn$ kernel~\eqref{eq:Matern} with parameter $\nu=5/2$  is adopted for the kernel $\kappa_x$ over continuous variables.
The weight $\lambda$ regarding the sum and product kernel compositions in \eqref{cx} is set to 0.5. The coefficients $\zeta_t=\hat{\zeta}_t=2,\forall t,$ in UCB-based acquisition rules \eqref{acq_problem} and \eqref{acq_problem_context}. Unless otherwise stated, the other kernel hyperparameters are optimized by maximizing the log marginal likelihood every $\delta=10$ slots via multi-started gradient descent. The performance measure of the competing methods is given by the notion of regret. By denoting the maximizer of $\varphi_t$ as $(\mathbf{c}^*_t,\mathbf{x}^*_t)$, the instantaneous regret per slot $t$ is $g_t:=\varphi_t(\mathbf{c}^*_t,\mathbf{x}^*_t)-\varphi_t(\mathbf{c}_t,\mathbf{x}_t)$, based on which the cumulative and average regrets are denoted as $G_T:=\sum_{t=1}^Tg_t$ and $\bar{G}_T:=G_T/T$, respectively. It is worth mentioning that $(\mathbf{c}^*_t,\mathbf{x}^*_t)$ are obtained by relying on explicit cost function in (P2) with known system state information. All the methods are run for 200 time slots and the average performances over 100 random repetitions are reported.

\subsection{Effect of Kernel Hyperparameters }

To study the effect of temporal and contextual kernel hyperparameters on the performance of the proposed BO approaches, a 2-BS MEC system with $M=2$ WDs is first considered, where the distances from the WDs to BSs are $[d_{1,1},d_{1,2},d_{2,1},d_{2,2}]=[20, 13, 15, 18]$ meters, the Rician factor in \eqref{eq: h_t} used to generate the channel gain is $K=4$, and $\eta$ in~\eqref{eq: state_gen} is set to $0.2$. Fig.~2 depicts the average regret of time-varying BO as a function of the time slot under different values of the temporal kernel hyperparameter $\rho$ in \eqref{temp}. It can be readily observed that the regret performance improves and then deteriorates as the value of $\rho$ increases. Specifically, $\rho=0.048$ achieves the lowest average regret by best capturing the temporal variation in the black-box objective function.

%that a too small (e.g., 0.04) or a too large (e.g., 0.05) temporal kernel parameter $\rho$ leads to a worse regret performance, where $\rho=0.048$ achieves the lowest average regret by adequately capturing the time variation of the black-box objective function.

\begin{figure}[htb]
\begin{centering}
\includegraphics[scale=0.8]{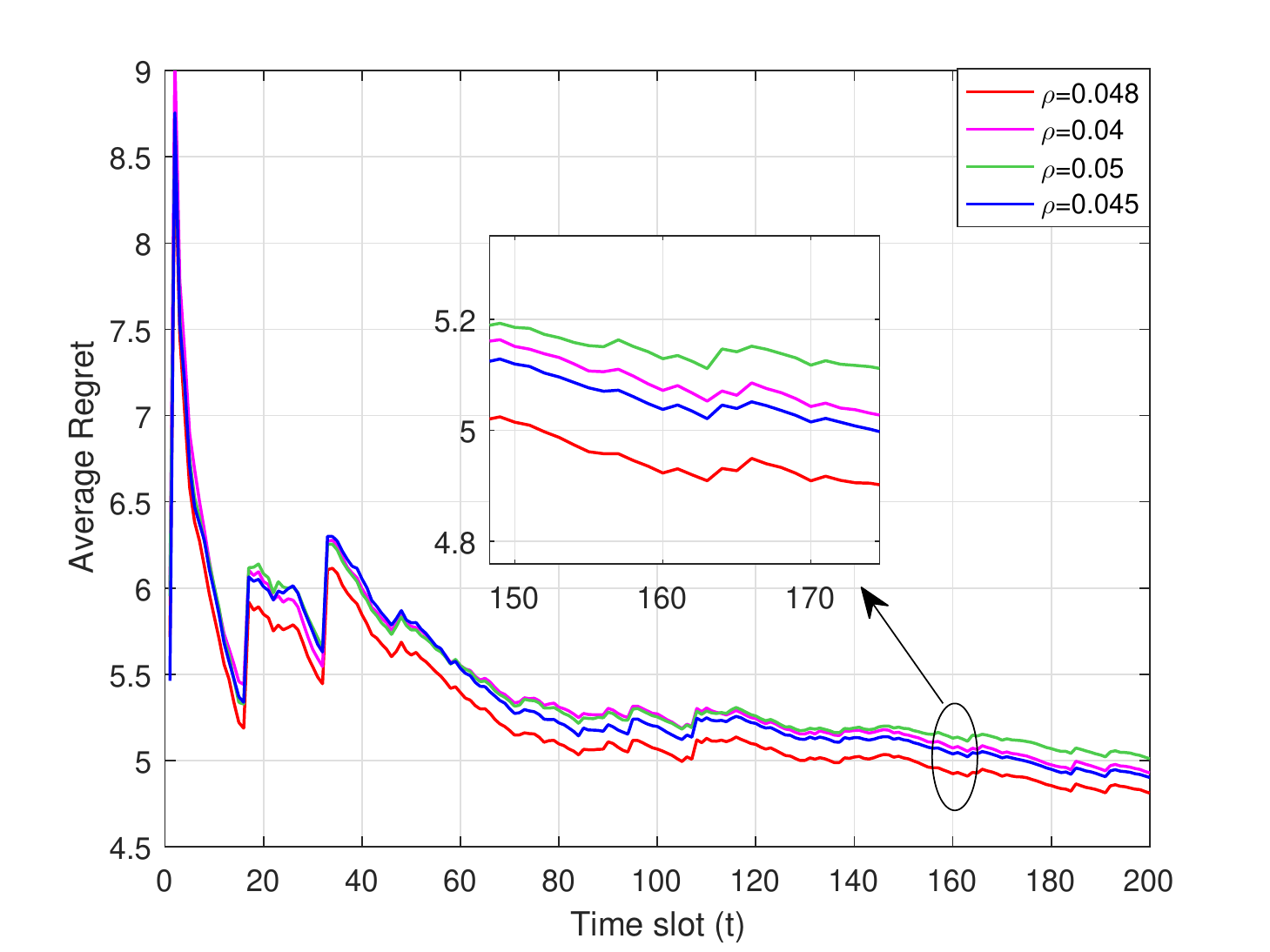}
\vspace{-0.1cm}
 \caption{Average regret under different temporal kernel hyperparameters $\rho$ for the time-varying BO approach.}
\end{centering}
\vspace{-0.1cm}
\end{figure}

Further considering the contextual time-varying BO where a $Mat\acute{e}rn$ kernel with $\nu = 5/2$ in \eqref{eq:Matern} is adopted for the contextual $\kappa_{s}$, the curves of the average regret for various contextual and temporal kernel hyperparameters are presented in Fig.~3, where it is evident that the best-performing
hyperparameter set is given by $\rho=0.02$ and $l=0.2$ in the temporal and contextual kernel, respectively.
%In Fig. 3, the average regrets are compared among different contextual and temporal kernel hyperparameters for the contextual time-varying BO algorithm, where a $Mat\acute{e}rn52$ kernel is used for the contextual kernel $\kappa_{con}$. It is observed that the contextual time-varying BO algorithm shows the best performance by properly choosing the temporal kernel parameter $\rho=0.02$ and the length-scale parameter $l=0.2$ in contextual kernel.
Notice that the best-performing hyperparameter $\rho$ of the temporal kernel in the contextual time-varying BO is smaller than that in the time-varying BO. To put it equivalently, the temporal kernel in the latter captures more dynamics in the objective function than that in the former. This phenomenon can be explained by that the observed contextual state information including time-varying task computational workload $L_t^m$ and input data size $I_t^m$ accounts for a portion of the overall dynamics, yielding lower degree of dynamics to be represented by the temporal kernel in the contextual time-varying BO.

%the correlation of partial system dynamics $\mathbf{s}_t$ including task computational workload and input data size is represented by the contextual kernel $\kappa_{con}$, while the temporal kernel $\kappa_{temp}$ characterizes the covariance regarding the other unknown system state information (e.g., wireless channel conditions and computing capabilities of edge servers). Compared with the time-varying BO algorithm, the temporal kernel parameter $\rho$ in the contextual time-varying BO algorithm is smaller due to the partially observed system dynamics encoded by the contextual kernel.

\begin{figure}[htb]
\begin{centering}
\includegraphics[scale=0.8]{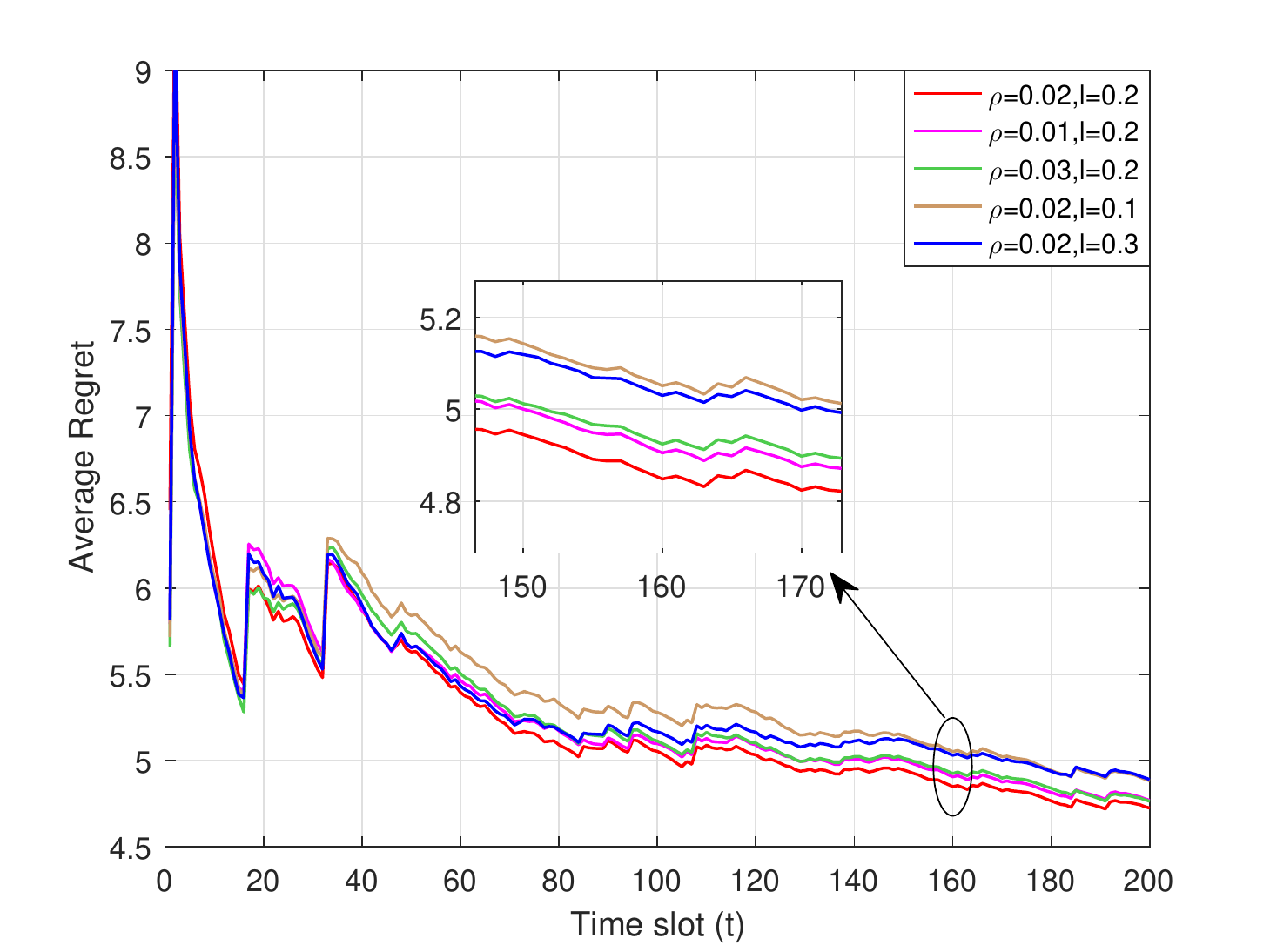}
\vspace{-0.1cm}
 \caption{Average regret under different contextual and temporal kernel hyperparameters for the contextual time-varying BO approach. }
\end{centering}
\vspace{-0.1cm}
\end{figure}

\subsection{Performance Comparison}

For performance comparison, three existing schemes are employed as baselines, namely, the MAB \cite{EXP3}, bandit convex optimization (BCO) \cite{BOC1}, and the conventional time-invariant BO approach~\cite{BO_tuto}. Since MAB can only cope with discrete decision variables, we discretized the analog-amplitude resource allocation variables into 5 levels and then adopted the multi-agent EXP3 method \cite{EXP3} for learning.  In BCO, the analog-amplitude resource allocation variables are obtained by constructing gradient estimates using evaluated function values, while the discrete offloading variables are still sought based on MAB as in the proposed BO approaches. Besides, time-invariant BO method neglects both temporal and contextual information in MEC systems.

\begin{figure}[htb]
\begin{centering}
\includegraphics[scale=0.8]{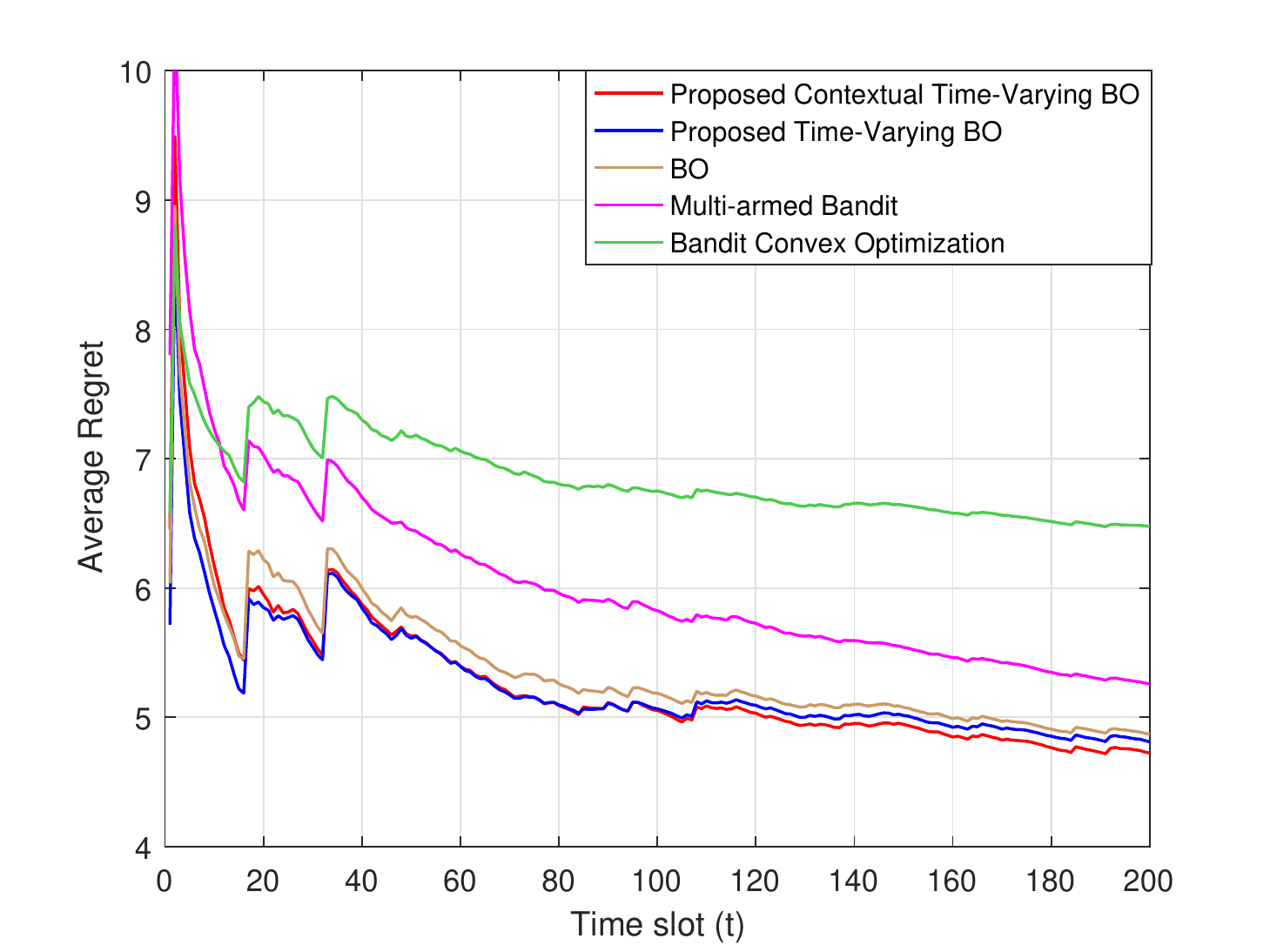}
\vspace{-0.1cm}
 \caption{Comparison of average regret under the 2-BS and 2-WD MEC system with Rician factor $K=4$ and $\eta=0.2$.}
\end{centering}
\vspace{-0.1cm}
\end{figure}

With properly selected temporal and contextual kernel hyperparameters, the average regret curves of all the competing approaches are presented in Fig. 4 for the 2-BS and 2-WD MEC system with $[d_{1,1},d_{1,2},d_{2,1},d_{2,2}]=[20, 13, 15, 18]$, $K=4$ and $\eta=0.2$. Specifically, the temporal kernel hyperparameter in the time-varying BO approach is chosen as $\rho=0.048$. As for contextual time-varying BO algorithm, the temporal kernel hyperparameter $\rho$ and the lengthscale $l$ of the contextual kernel are set to $0.02$ and $0.2$, respectively.
As shown in Fig.~4,
our proposed time-varying BO approach outperforms the three benchmarks, namely, time-invariant BO, MAB, and BCO, by around $1.21\%$, $8.51\%$ and $25.72\%$ in average regret after $200$ time slots. This suggests the benefits of adapting temporal information-aided Bayesian approach to the black-box optimization with both categorical (i.e., task offloading) and analog-amplitude (i.e., resource allocation) variables. By further utilizing the observed context information (i.e., the characteristics of computational tasks) via the contextual kernel, the novel contextual time-varying BO method achieves $1.81\%$ and $3\%$ lower average regret than time-varying BO and traditional BO after 200 slots.

\begin{figure}[htb]
\begin{centering}
\includegraphics[scale=0.8]{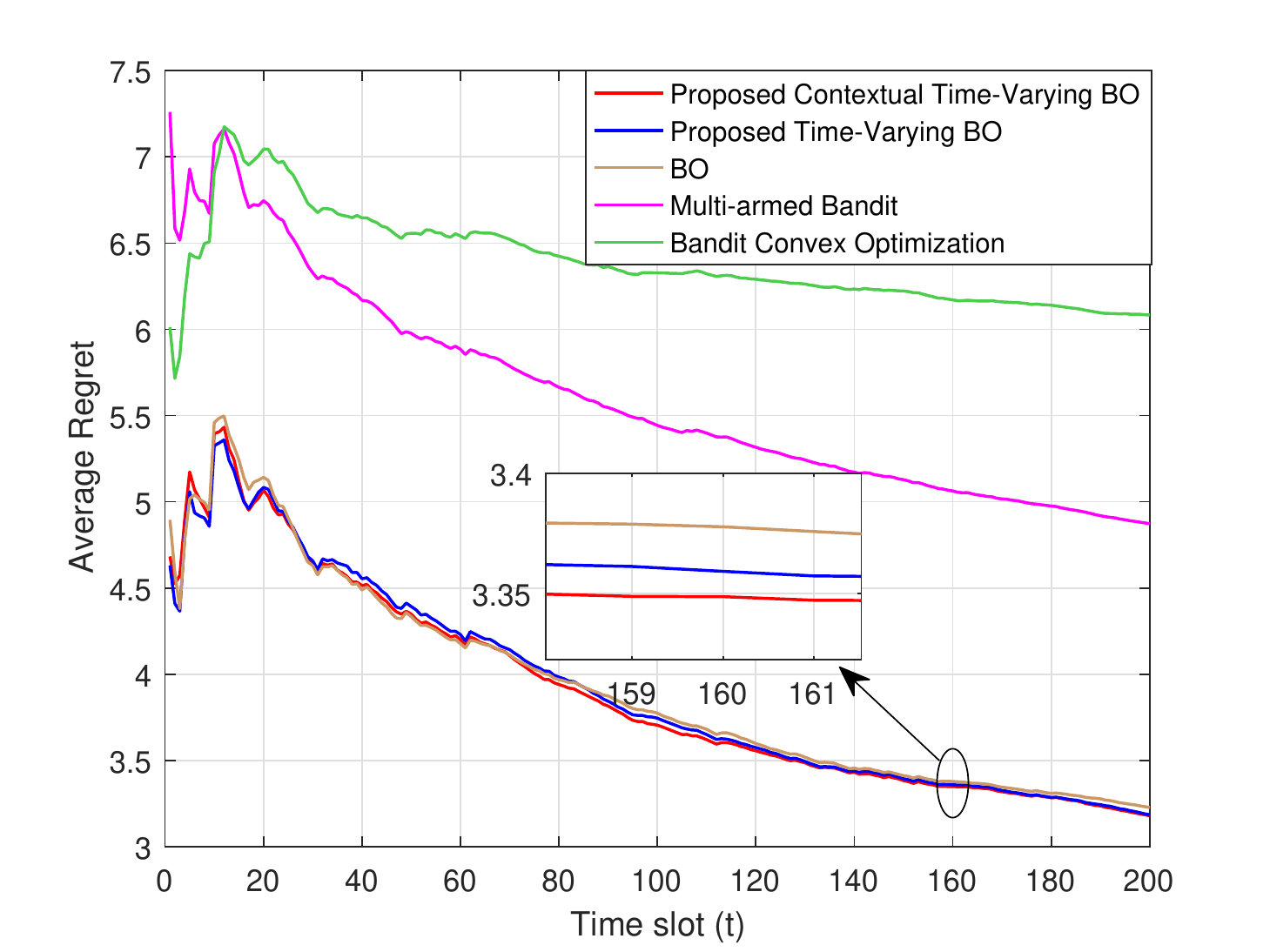}
\vspace{-0.1cm}
 \caption{Comparison of average regret under the 2-BS and 2-WD MEC system with Rician factor $K=9$ and $\eta=0.02$.}
\end{centering}
\vspace{-0.1cm}
\end{figure}

Further, the performances of the proposed BO approaches are investigated in the $2$-BS and $2$-WD MEC system with a smaller scale of system dynamics, that is given by the Rician factor $K=9$ in~\eqref{eq: h_t} and the temporal variation parameter $\eta=0.02$ in~\eqref{eq: state_gen}.
%How our proposed BO algorithms perform in small-variation system dynamics is shown in Fig. 5, where the distances between the 2 BSs and 2 WDs are $[d_{1,1},d_{1,2},d_{2,1},d_{2,2}]=[20, 13, 15, 18]$, the Rician factor $\theta=0.9$ and the time variation indicator $\eta=0.02$.
The temporal kernel parameter $\rho$ is set to $0.011$ in the time-varying BO approach, while $\rho=0.0045$ and contextual kernel lengthscale $l=0.2$ are chosen in contextual time-varying BO. Here, the values of $\rho$ in both cases are smaller than the counterparts in Fig.~4, what is in accordance with the degree of the underlying temporal dynamics. Compared with the alternative time-invariant BO, MAB, and BCO schemes, the proposed (contextual) time-varying BO methods reduce the average regret by approximately $1.49\%$, $34.77\%$ and $47.75\%$ after $200$ slots as showcased in Fig.~5. In addition, the performance of the time-invariant BO method is close to the proposed time-varying BO alternatives due to such small-scale system dynamics.

%\changeQ{It can be seen that the proposed BO algorithms outperform the three benchmarks, e.g., around $1.49\%$, $34.77\%$ and $47.75\%$ average regret improvement after $200$ time slots for contextual time-varying BO, compared with time-invariant BO, MAB and BCO schemes, respectively. Aiming at reflecting the slow time variation, the temporal kernel parameters $\rho$ are smaller than that in Fig. 4 for both time-varying BO and contextual time-varying BO approaches. In addition, the performance of the time-invariant BO method is close to the proposed time-varying BO algorithm under such small-variation system dynamics.}

\subsection{Effect of Network Size}

\begin{figure}[htb]
\begin{centering}
\includegraphics[scale=0.8]{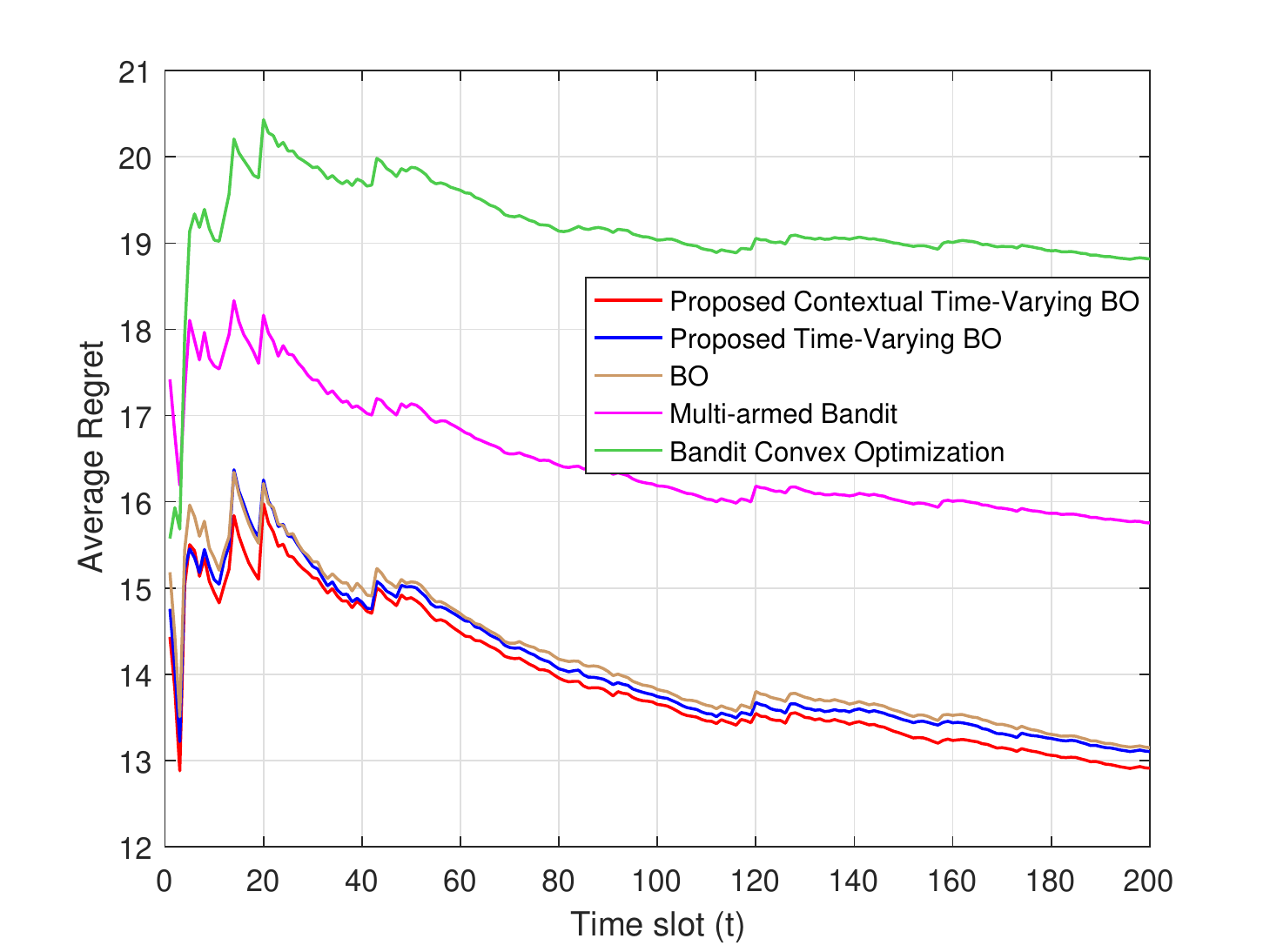}
\vspace{-0.1cm}
 \caption{Comparison of average regret under the 2-BS and 5-WD MEC system with Rician factor $K=5.67$ and $\eta=0.2$.}
\end{centering}
\vspace{-0.1cm}
\end{figure}

Lastly, the performances of all the schemes are assessed as the number of WDs and BSs varies. Consider first a 2-BS MEC system with a larger number $M=5$ of WDs, where the time-varying system state is generated using the Rician factor $K=5.67$~in \eqref{eq: h_t} and temporal variation factor $\eta=0.2$ in~\eqref{eq: state_gen}. In this case, $\rho=0.018$ in time-varying BO approach, while $\rho=0.006$ and $l=0.5$ in contextual time-varying BO strategy. Still, the proposed (contextual) time-varying BO methods outperform the other three alternatives by leveraging temporal and contextual information as shown in Fig.~6.
%The tests in this subsection evaluate the performance of all schemes under different number of WDs and BSs (i.e., network size), where the distances between BSs and WDs are randomly generated within the range $[5,20]$. The average regrets are first compared among the proposed BO algorithms and benchmark methods in Fig. 6 under the 2-BS and 5-WD MEC system with Rician factor $\theta=0.85$ and time variation indicator $\eta=0.2$. In this case, $\rho=0.018$ in time-varying BO approach, while $\rho=0.006$ and $l=0.5$ in contextual time-varying BO strategy. Clearly, by leveraging temporal and contextual information, the proposed contextual time-varying BO and time-varying BO algorithms enjoy lower average regrets than traditional BO, MAB and BCO schemes.

Moreover, fixing the number $M$ of WDs as $2$, the average EDC over slots is plotted as a function of the number $N$ of BSs for all the competing methods in Fig.~7. Here, the Rician factor~in \eqref{eq: h_t} and value of $\eta$ in~\eqref{eq: state_gen} are set to $K=4$ and $\eta=0.2$ respectively. Apparently, the two proposed BO approaches achieve lower average EDC than the other three baselines. Additionally, the average EDC of all the methods decreases as the network size grows by better exploiting the diverse computing capacities and channel conditions of the edge servers.

%the EDC averaged over the time under different number of edge servers is presented in Fig. 7, where the number of WDs $M=2$, and the Rician factor and time variation indicator are set to $\theta=0.8$ and $\eta=0.2$ respectively. Under different network sizes, the proposed BO algorithms achieve lower average EDC than the three representative benchmarks. Additionally, the average EDC decreases as the network size grows by exploiting the heterogeneity of more edge servers.

\begin{figure}[htb]
\begin{centering}
\includegraphics[scale=0.8]{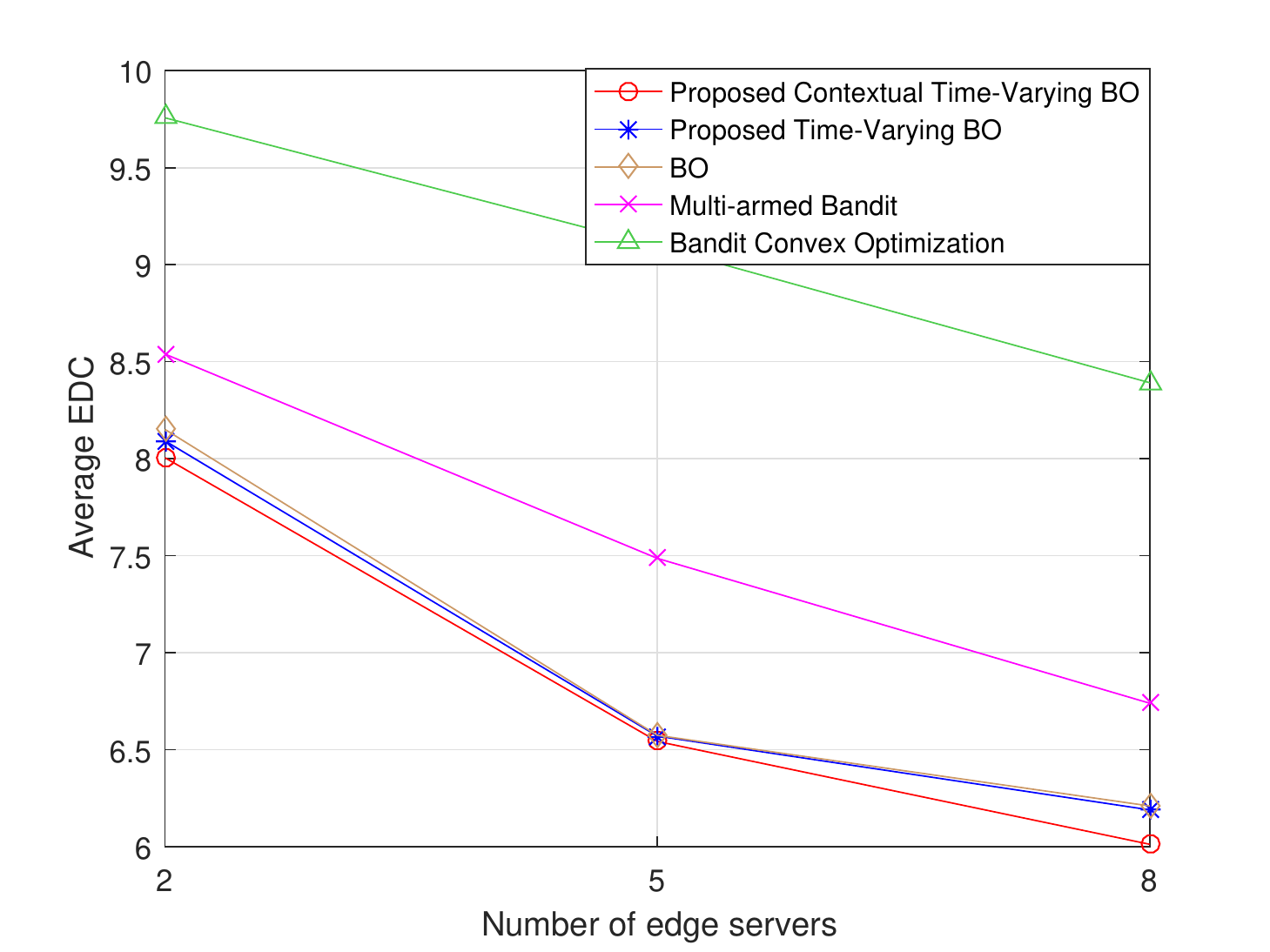}
\vspace{-0.1cm}
 \caption{Impact of MEC network size on average energy-delay cost.}
\end{centering}
\vspace{-0.1cm}
\end{figure}

\section{Conclusion}

BO for dynamic MEC management was studied in this paper. Different from prior works in time-varying MEC systems, the focus was online joint optimization of discrete task offloading decisions and analog-amplitude resource allocation strategies by minimizing the EDC using only bandit observations at queried points. Specifically, by exploiting both temporal and contextual information, we developed two novel BO approaches that incorporate the strength of the MAB framework. Numerical tests under different MEC network sizes demonstrated the effectiveness of the proposed BO approaches.

\begin{footnotesize}
\bibliographystyle{IEEEtran}

\begin{thebibliography}{10}
\providecommand{\url}[1]{#1}
\csname url@samestyle\endcsname
\providecommand{\newblock}{\relax}
\providecommand{\bibinfo}[2]{#2}
\providecommand{\BIBentrySTDinterwordspacing}{\spaceskip=0pt\relax}
\providecommand{\BIBentryALTinterwordstretchfactor}{4}
\providecommand{\BIBentryALTinterwordspacing}{\spaceskip=\fontdimen2\font plus
\BIBentryALTinterwordstretchfactor\fontdimen3\font minus
  \fontdimen4\font\relax}
\providecommand{\BIBforeignlanguage}[2]{{%
\expandafter\ifx\csname l@#1\endcsname\relax
\typeout{** WARNING: IEEEtran.bst: No hyphenation pattern has been}%
\typeout{** loaded for the language `#1'. Using the pattern for}%
\typeout{** the default language instead.}%
\else
\language=\csname l@#1\endcsname
\fi
#2}}
\providecommand{\BIBdecl}{\relax}
\BIBdecl

\bibitem{jiaasilomar}
J.~Yan, Q.~Lu, and G.~B. Giannakis, ``Bayesian optimization for task offloading
  and resource allocation in mobile edge computing,'' \emph{Proc. Asilomar
  Conf. Sig., Syst., Comput.}, 2022.

\bibitem{MECsurvey1}
Y.~Mao, C.~You, J.~Zhang, K.~Huang, and K.~B. Letaief, ``A survey on mobile
  edge computing: The communication perspective,'' \emph{IEEE Commun. Surveys
  Tuts.}, vol.~19, no.~4, pp. 2322--2358, Fourthquarter 2017.

\bibitem{MEC2}
C.~You, K.~Huang, and H.~Chae, ``Energy efficient mobile cloud computing
  powered by wireless energy transfer,'' \emph{{IEEE} J. Sel. Areas Commun.},
  vol.~34, no.~5, pp. 1757--1771, May 2016.

\bibitem{MEC1}
C.~You, K.~Huang, H.~Chae, and B.~H. Kim, ``Energy-efficient resource
  allocation for mobile-edge computation offloading,'' \emph{{IEEE} Trans.
  Wireless Commun.}, vol.~16, no.~3, pp. 1397--1411, Mar. 2017.

\bibitem{wang2016mobile}
Y.~Wang, M.~Sheng, X.~Wang, L.~Wang, and J.~Li, ``Mobile-edge computing:
  {P}artial computation offloading using dynamic voltage scaling,''
  \emph{{IEEE} Trans. Commun.}, vol.~64, no.~10, pp. 4268--4282, 2016.

\bibitem{MEC3}
T.~Q. Dinh, J.~Tang, Q.~D. La, and T.~Q.~S. Quek, ``Offloading in mobile edge
  computing: Task allocation and computational frequency scaling,''
  \emph{{IEEE} Trans. Commun.}, vol.~65, no.~8, pp. 3571--3584, 2017.

\bibitem{bi2018computation}
S.~Bi and Y.~J. Zhang, ``Computation rate maximization for wireless powered
  mobile-edge computing with binary computation offloading,'' \emph{{IEEE}
  Trans. Wireless Commun.}, vol.~17, no.~6, pp. 4177--4190, 2018.

\bibitem{yan2019optimal}
J.~Yan, S.~Bi, Y.~J. Zhang, and M.~Tao, ``Optimal task offloading and resource
  allocation in mobile-edge computing with inter-user task dependency,''
  \emph{{IEEE} Trans. Wireless Commun.}, vol.~19, no.~1, pp. 235--250, 2019.

\bibitem{mao2016dynamic}
Y.~Mao, J.~Zhang, and K.~B. Letaief, ``Dynamic computation offloading for
  mobile-edge computing with energy harvesting devices,'' \emph{{IEEE} J. Sel.
  Areas Commun.}, vol.~34, no.~12, pp. 3590--3605, 2016.

\bibitem{mao2017stochastic}
Y.~Mao, J.~Zhang, S.~Song, and K.~B. Letaief, ``Stochastic joint radio and
  computational resource management for multi-user mobile-edge computing
  systems,'' \emph{{IEEE} Trans. Wireless Commun.}, vol.~16, no.~9, pp.
  5994--6009, 2017.

\bibitem{yang2022dynamic}
Z.~Yang, S.~Bi, and Y.-J.~A. Zhang, ``Dynamic offloading and trajectory control
  for {UAV}-enabled mobile edge computing system with energy harvesting
  devices,'' \emph{{IEEE} Trans. Wireless Commun.}, 2022.

\bibitem{chen2018heterogeneous}
T.~Chen, Q.~Ling, Y.~Shen, and G.~B. Giannakis, ``Heterogeneous online learning
  for ``thing-adaptive" fog computing in {IoT},'' \emph{IEEE Internet of Things
  Journal}, vol.~5, no.~6, pp. 4328--4341, 2018.

\bibitem{chen2017online}
T.~Chen, Q.~Ling, and G.~B. Giannakis, ``An online convex optimization approach
  to proactive network resource allocation,'' \emph{{IEEE} Trans. Signal
  Process.}, vol.~65, no.~24, pp. 6350--6364, 2017.

\bibitem{hall2015online}
E.~C. Hall and R.~M. Willett, ``Online convex optimization in dynamic
  environments,'' \emph{{IEEE} J. Sel. Topics Signal Process.}, vol.~9, no.~4,
  pp. 647--662, 2015.

\bibitem{BOC1}
A.~D. Flaxman, A.~T. Kalai, and H.~B. McMahan, ``Online convex optimization in
  the bandit setting: Gradient descent without a gradient,'' in \emph{Proc. ACM
  SODA, Vancouver, BC, Canada}, Jan. 2005, pp. 385--394.

\bibitem{agarwal2010optimal}
A.~Agarwal, O.~Dekel, and L.~Xiao, ``Optimal algorithms for online convex
  optimization with multi-point bandit feedback.'' \emph{Proc. Annual Conf.
  Learning Theory}, pp. 28--40, 2010.

\bibitem{shamir2017optimal}
O.~Shamir, ``An optimal algorithm for bandit and zero-order convex optimization
  with two-point feedback,'' \emph{J. Mach. Learn. Res.}, vol.~18, no.~1, pp.
  1703--1713, 2017.

\bibitem{BOC2}
T.~Chen and G.~B. Giannakis, ``Bandit convex optimization for scalable and
  dynamic {IoT} management,'' \emph{IEEE Internet of Things Journal}, vol.~6,
  no.~1, pp. 1276--1286, 2019.

\bibitem{MEC_MAB1}
B.~Wu, T.~Chen, W.~Ni, and X.~Wang, ``Multi-agent multi-armed bandit learning
  for online management of edge-assisted computing,'' \emph{{IEEE} Trans.
  Commun.}, vol.~69, no.~12, pp. 8188--8199, 2021.

\bibitem{MEC_MAB2}
B.~Li, T.~Chen, and G.~B. Giannakis, ``Secure mobile edge computing in {IoT}
  via collaborative online learning,'' \emph{{IEEE} Trans. Signal Process.},
  vol.~67, no.~23, pp. 5922--5935, 2019.

\bibitem{MEC_MAB3}
Y.~Sun, X.~Guo, J.~Song, S.~Zhou, Z.~Jiang, X.~Liu, and Z.~Niu, ``Adaptive
  learning-based task offloading for vehicular edge computing systems,''
  \emph{{IEEE} Trans. Veh. Technol.}, vol.~68, no.~4, pp. 3061--3074, 2019.

\bibitem{sun2017emm}
Y.~Sun, S.~Zhou, and J.~Xu, ``{EMM}: Energy-aware mobility management for
  mobile edge computing in ultra dense networks,'' \emph{{IEEE} J. Sel. Areas
  Commun.}, vol.~35, no.~11, pp. 2637--2646, 2017.

\bibitem{BO_tuto}
P.~I. Frazier, ``A tutorial on {B}ayesian optimization,''
  \emph{arXiv:1807.02811. [Online]. Available:
  http://arxiv.org/abs/1807.02811}, 2018.

\bibitem{lu2022surrogate}
Q.~Lu, K.~D. Polyzos, B.~Li, and G.~B. Giannakis, ``Surrogate modeling for
  {B}ayesian optimization beyond a single {G}aussian process,'' \emph{arXiv
  preprint arXiv:2205.14090}, 2022.

\bibitem{Rasmussen2006gaussian}
C.~E. Rasmussen and C.~K. Williams, \emph{Gaussian processes for Machine
  Learning}.\hskip 1em plus 0.5em minus 0.4em\relax MIT press Cambridge, MA,
  2006.

\bibitem{lu2020ensemble}
Q.~Lu, G.~Karanikolas, Y.~Shen, and G.~B. Giannakis, ``Ensemble {G}aussian
  processes with spectral features for online interactive learning with
  scalability,'' \emph{Proc. Int. Conf. Artif. Intel. and Stats.}, pp.
  1910--1920, 2020.

\bibitem{lu2022incremental}
Q.~Lu, G.~V. Karanikolas, and G.~B. Giannakis, ``Incremental ensemble
  {G}aussian processes,'' \emph{IEEE Trans. Pattern Anal. Mach. Intel.}, 2022.

\bibitem{polyzos2021ensemble}
K.~D. Polyzos, Q.~Lu, and G.~B. Giannakis, ``Ensemble {G}aussian processes for
  online learning over graphs with adaptivity and scalability,'' \emph{IEEE
  Trans. Sig. Process.}, 2021.

\bibitem{snoek2012practical}
J.~Snoek, H.~Larochelle, and R.~P. Adams, ``Practical {B}ayesian optimization
  of machine learning algorithms,'' \emph{Proc. Adv. Neural Inf. Process.
  Syst.}, vol.~25, 2012.

\bibitem{korovina2020chembo}
K.~Korovina, S.~Xu, K.~Kandasamy, W.~Neiswanger, B.~Poczos, J.~Schneider, and
  E.~Xing, ``Chembo: {B}ayesian optimization of small organic molecules with
  synthesizable recommendations,'' \emph{Proc. Int. Conf. Artif. Intel. and
  Stats.}, pp. 3393--3403, 2020.

\bibitem{cully2015robots}
A.~Cully, J.~Clune, D.~Tarapore, and J.-B. Mouret, ``Robots that can adapt like
  animals,'' \emph{Nature}, vol. 521, no. 7553, pp. 503--507, 2015.

\bibitem{BO_RA}
L.~Maggi, A.~Valcarce, and J.~Hoydis, ``Bayesian optimization for radio
  resource management: Open loop power control,'' \emph{{IEEE} J. Sel. Areas
  Commun.}, vol.~39, no.~7, pp. 1858--1871, 2021.

\bibitem{dreifuerst2021optimizing}
R.~M. Dreifuerst, S.~Daulton, Y.~Qian, P.~Varkey, M.~Balandat, S.~Kasturia,
  A.~Tomar, A.~Yazdan, V.~Ponnampalam, and R.~W. Heath, ``Optimizing coverage
  and capacity in cellular networks using machine learning,'' \emph{Proc. IEEE
  Int. Conf. Acoust., Speech, Sig. Process.}, pp. 8138--8142, 2021.

\bibitem{yang2022bayesian}
S.~Yang, B.~Liu, Z.~Hong, and Z.~Zhang, ``Bayesian optimization-based beam
  alignment for {MmWave MIMO} communication systems,'' \emph{arXiv preprint
  arXiv:2207.14174}, 2022.

\bibitem{liu2022deep}
H.~Liu and G.~Cao, ``Deep learning video analytics through online learning
  based edge computing,'' \emph{{IEEE} Trans. Wireless Commun.}, 2022.

\bibitem{cocabo}
B.~Ru, A.~S. Alvi, V.~Nguyen, M.~A. Osborne, and S.~J. Roberts, ``Bayesian
  optimisation over multiple continuous and categorical inputs,'' \emph{Proc.
  Int. Conf. Mach. Learn.}, 2020.

\bibitem{bogunovic2016time}
I.~Bogunovic, J.~Scarlett, and V.~Cevher, ``Time-varying {G}aussian process
  bandit optimization,'' \emph{Proc. Int. Conf. Artif. Intel. and Stats.}, pp.
  314--323, 2016.

\bibitem{EXP3}
A.~Peter, C.-B. Nicolo, F.~Yoav, and R.~E. Schapire, ``The nonstochastic
  multiarmed bandit problem,'' \emph{SIAM J. on Computing}, pp. 48--77, 2002b.

\bibitem{srinivas2012information}
N.~Srinivas, A.~Krause, S.~M. Kakade, and M.~W. Seeger, ``Information-theoretic
  regret bounds for {G}aussian process optimization in the bandit setting,''
  \emph{IEEE Trans. Inf. Theory}, vol.~58, no.~5, pp. 3250--3265, 2012.

\bibitem{krause2011contextual}
A.~Krause and C.~Ong, ``Contextual {G}aussian process bandit optimization,''
  \emph{Proc. Adv. Neural Inf. Process. Syst.}, vol.~24, 2011.

\bibitem{miettinen2010energy}
A.~P. Miettinen and J.~K. Nurminen, ``Energy efficiency of mobile clients in
  cloud computing,'' in \emph{2nd USENIX Workshop on Hot Topics in Cloud
  Computing (HotCloud 10)}, 2010.

\end{thebibliography}

\end{footnotesize}

\end{document}